\documentclass[namedreferences]{solarphysics}
\usepackage[hyperref,optionalrh,solaromanenum]{spr-sola-addons} 
\usepackage{graphicx}                    
\usepackage{color}                       
\usepackage{breakurl}                         

\usepackage[dvipsnames]{xcolor}   

\begin{document}



\newcommand{\degree}{^\circ} 
\newcommand{\kms}{km s$^{-1}$} 
\newcommand{\Rsun}{$R_{\odot}$}  

\newcommand{\f}[2]{{\ensuremath{\mathchoice%
        {\dfrac{#1}{#2}}
        {\dfrac{#1}{#2}}
        {\frac{#1}{#2}}
        {\frac{#1}{#2}}
        }}}
\newcommand{\Int}[2]{\ensuremath{\mathchoice%
        {\displaystyle\int_{#1}^{#2}}
        {\displaystyle\int_{#1}^{#2}}
        {\int_{#1}^{#2}}
        {\int_{#1}^{#2}}
        }}
\newcommand{\curl}{ {\bf \nabla } \times}
\newcommand{\rmd}{{\rm d }}
\renewcommand{\div}[1]{ {\bf \nabla}. #1 }
\newcommand{\grad}{  {\bf \nabla} }
\newcommand{\pder}[2]{\f{\partial #1}{\partial #2}}
\newcommand{\der}[2]{\f{\rmd \, #1}{\rmd \, #2}}

\newcommand{\BE}{\begin{equation}}
\newcommand{\EE}{\end{equation}}
\newcommand{\BA}{\begin{eqnarray}}
\newcommand{\EA}{\end{eqnarray}}
\newcommand{\Fig}[1]{Figure~\ref{fig:#1}}
\newcommand{\fig}[1]{Figure~\ref{fig:#1}}
\newcommand{\figsss}[1]{Figure~\ref{fig:#1}}
\newcommand{\figs}[2]{Figures~\ref{fig:#1} and \ref{fig:#2}}
\newcommand{\figss}[2]{Figures.~\ref{fig:#1} - \ref{fig:#2}}
\newcommand{\sect}[1]{Section~\ref{s:#1}}
\newcommand{\app}[1]{Appendix~\ref{app_#1}}
\newcommand{\sects}[2]{Sects.~\ref{sect_#1} and~\ref{sect_#2}}
\newcommand{\eq}[1]{Equation~\ref{eq_#1}}
\newcommand{\eqs}[2]{Eqs.~\ref{eq_#1} and \ref{eq_#2}}
\newcommand{\eqss}[2]{Eqs.~\ref{eq_#1} - \ref{eq_#2}}
\newcommand{\eqsss}[3]{Eqs.~\ref{eq_#1}, \ref{eq_#2} and \ref{eq_#3}}
\newcommand{\eg}{{ e.g.}}
\newcommand{\etal}{{\it et al.}}
\newcommand{\ie}{{\it i.e.}}
\newcommand{\insitu}{{\it in situ}}
\newcommand{\cf}{{\it cf.}}

\newcommand{\Bo}{B_{\rm 0}}
\newcommand{\dnot}{d_{\rm 0}}
\newcommand{\Nt}{N_{\rm t}}
\newcommand{\yc}{y_{\rm c}}
\newcommand{\xc}{x_{\rm c}}
\newcommand{\minn}{{\rm min}}
\newcommand{\maxx}{{\rm max}}
\newcommand{\phic}{\phi_{\rm c}}
\newcommand{\tauc}{\tau_{\rm c}}
\newcommand{\Fmax}{F_{\rm max}}
\newcommand{\Bz}{B_{\rm z}}
\newcommand{\Cit}{C_{\rm it}}
\newcommand{\Fz}{F_{\rm z}}
\newcommand{\Fzax}{F_{\rm z}^{\rm axial}}
\newcommand{\Fzaz}{F_{\rm z}^{\rm azimuthal}}
\newcommand{\nit}{n_{\rm it}}
\newcommand{\phia}{\phi^{\rm M}_{\rm a}}
\newcommand{\phii}{\phi_{\rm i}}
\newcommand{\phiwl}{\phi^{\rm WL}_{\rm U}}
\newcommand{\phiwlm}{\phi^{\rm WLM}_{\rm U}}
\newcommand{\comment}[1]{}

\newcommand{\Bj}[1]{B_{#1,j}}
\newcommand{\cL}{\mathcal{L}}
\newcommand{\cP}{\mathcal{P}}
\newcommand{\vNabla}{\vec{\nabla}_p}
\newcommand{\vp}{\vec{\mathfrak{p}}}
\newcommand{\vpb}{\vec{\mathfrak{p}_b}}
\newcommand{\Mp}{M_{\vp}}
\newcommand{\Mpi}{M_{\vp_i}}
\newcommand{\Mo}{M_o}
\newcommand{\Mr}{M_r}
\newcommand{\np}{n_p}

\newcommand{\adv}{    {\it Adv. Space Res.}} 
\newcommand{\annG}{   {\it Ann. Geophys.}} 
\newcommand{\aap}{    {\it Astron. Astrophys.}}
\newcommand{\aaps}{   {\it Astron. Astrophys. Suppl.}}
\newcommand{\aapr}{   {\it Astron. Astrophys. Rev.}}
\newcommand{\ag}{     {\it Ann. Geophys.}}
\newcommand{\aj}{     {\it Astron. J.}} 
\newcommand{\apj}{    {\it Astrophys. J.}}
\newcommand{\apjs}{    {\it Astrophys. J. Sup.}}
\newcommand{\apjl}{   {\it Astrophys. J. Lett.}}
\newcommand{\apss}{   {\it Astrophys. Space Sci.}} 
\newcommand{\cjaa}{   {\it Chin. J. Astron. Astrophys.}} 
\newcommand{\gafd}{   {\it Geophys. Astrophys. Fluid Dyn.}}
\newcommand{\grl}{    {\it Geophys. Res. Lett.}}
\newcommand{\ijga}{   {\it Int. J. Geomagn. Aeron.}}
\newcommand{\jastp}{  {\it J. Atmos. Solar-Terr. Phys.}} 
\newcommand{\jgr}{    {\it J. Geophys. Res.}}
\newcommand{\mnras}{  {\it Mon. Not. Roy. Astron. Soc.}}
\newcommand{\nat}{    {\it Nature}}
\newcommand{\pasp}{   {\it Pub. Astron. Soc. Pac.}}
\newcommand{\pasj}{   {\it Pub. Astron. Soc. Japan}}
\newcommand{\pre}{    {\it Phys. Rev. E}}
\newcommand{\physrep}{    {\it Phys. Rep.}}

\newcommand{\solphys}{{\it Solar Phys.}}
\newcommand{\sovast}{ {\it Soviet  Astron.}} 
\newcommand{\ssr}{    {\it Space Sci. Rev.}}

\begin{article}

\begin{opening}

\title{Modeling Global Magnetic Flux Emergence in Bipolar Solar Active Regions}

%

\author[addressref={1},email={mpoisson@iafe.uba.ar}]{\inits{M. }\fnm{Mariano }\lnm{Poisson}\orcid{http://orcid.org/0000-0002-4300-0954}}
\author[addressref={1}]{\inits{M. }\fnm{Marcelo }\lnm{L\'opez Fuentes}\orcid{http://orcid.org/0000-0001-8830-4022}}
\author[addressref={1}]{\inits{C. H. }\fnm{Cristina H. }\lnm{Mandrini}\orcid{http://orcid.org/0000-0001-9311-678X}}
\author[addressref={2,3}]{\inits{P.  }\fnm{Pascal }\lnm{D\'emoulin}\orcid{http://orcid.org/0000-0001-8215-6532}}
\author[addressref={1}]{\inits{F.  }\fnm{Francisco }\lnm{Grings}\orcid{http://orcid.org/0000-0001-5252-2466}}

%
\runningauthor{Poisson et al.}
\runningtitle{Modeling Magnetic Flux Emergence in Bipolar Active Regions}


\address[id={1}]{Instituto de Astronom\'{\i}a y F\'{\i}sica del Espacio (IAFE), CONICET-UBA, Ciudad Universitaria – Pab. 2, Intendente Güiraldes 2160,
C1428EGA C.A.B.A., Argentina}

\address[id={2}]{LESIA, Observatoire de Paris, Universit\'e PSL, CNRS, Sorbonne Universit\'e, Univ. Paris Diderot, Sorbonne Paris Cit\'e, 5 place Jules Janssen, 92195 Meudon, France.}

\address[id={3}]{Laboratoire Cogitamus, rue Descartes, 75005 Paris, France.}

\begin{abstract}
{%
Active regions (ARs) appear in the solar atmosphere as a consequence of the emergence of magnetic flux-ropes (FR). In this study, we use Bayesian methods to analyze line-of-sight magnetograms of emerging ARs. We employ a FR model consisting of a half-torus field structure based on eight parameters. The goal is to derive constrained physical parameters of the originating FR which are consistent with the observations. Specifically, we aim to obtain a precise estimation of the {AR} tilt angle and {magnetic} twist at different stages of the emergence {process}. To achieve this, {we propose} four temporal methods that correlate the {field parameter evolutions} with a single coherent FR. These methods differ from each other in the size of the explored parameter space. We test {the} methods on four bipolar ARs observed {with} the {\it Michelson Doppler Imager} on board the {\it Solar and Heliospheric Observatory.} 
We find that tilt angles are typically consistent between the temporal methods, improving previous estimations at all stages of the emergence. The twist sign derived from the temporal methods is consistent with previous estimations. 
The standard errors of all the methods used are similar, indicating that they model the observations equally well. These results indicate that the proposed methods can be used to obtain global magnetic parameters of ARs during their early evolution. The derived parameters contribute to a better understanding of the formation of FRs, and the role of ARs in the magnetic recycling process along the solar cycle.
}

\end{abstract}

%
\keywords{Active Regions, Magnetic Fields; Active Regions, Models; Magnetic fields, Photosphere}

\end{opening}

%

 \section{Introduction} \label{s:intro} 


Observations and analysis of active regions (ARs) along solar cycles are a fundamental ingredient of current dynamo models that aim to explain the processes of generation, amplification, and evolution of the solar magnetic field \citep[][]{Parker79,Brandenburg2005,Charbonneau13}. Several models locate the origin of ARs at the bottom of the convective zone (CZ), where the global poloidal magnetic field is transformed and accumulated into a toroidal component \citep[][and references therein]{Howe2009}. Simulations show that due to plasma instabilities in this region, magnetic field structures are formed as coherent flux tubes {that} travel throughout the convective zone due to the buoyant force {\citep{Spruit99,Miesch07,Nelson2013}}. 
{When} they finally emerge at the photospheric level their magnetic field expands {into} the corona {\citep{Archontis04, Gibson06}}. 
These flux tubes, {which are anchored at the bottom of the CZ, are known as $\Omega$ loops due to the shape they display} in different models. ARs are the manifestation of the emergence of these flux tubes into the solar atmosphere {\citep[][]{Fan09,Hood12,Cheung14}}. 
In particular, the line-of-sight (LOS) component of the photospheric magnetic field shows these $\Omega$ loops as bipolar magnetic regions.

Observational evidence of ARs indicates that the emerging flux ropes (FR) transport magnetic helicity \citep[][]{Gibson04}. 
The non-potential field configuration of ARs {relates to their level of activity}, since it is closely linked to the free magnetic energy stored and available to be released during energetic events such as flares {and} coronal mass ejections (CMEs) \citep[][]{Green02, Chandra09}.
Being magnetic helicity a well-conserved quantity in astrophysical plasmas \citep[][]{Berger84}, the estimation of this quantity at the Sun and in the interplanetary medium can be used as a proxy to correlate the {solar source of CMEs to their interplanetary counterparts} using in situ observations of magnetic clouds \citep[][]{Mandrini04,Dasso2009}. 

{
The helicity relative to the potential field configuration in {the emerging} FRs can be interpreted as the combination of two components. The twist, which is linked to the {number of} turns of the field lines around the FR axis, and the writhe, which describes the torsion of the FR main axis as a whole \citep[][]{Longcope98,Lopez-Fuentes03,Berger06}. This decomposition is useful for models and simulations in which the 3D magnetic field of the FR is known at a given moment, providing a measure of the contribution of each component to the total magnetic helicity. However, this decomposition might not be always practical, since it is not possible to identify or detach the contribution of each component from the observed photospheric magnetic field.}

{Numerical simulations have shown that emerging FRs need a certain amount of twist to maintain a cohesive structure during their transit through the CZ \citep{Emonet98}. But the estimations of this quantity on ARs may pose several observational limitations, so there is no actual consensus about the minimum amount of twist needed {to keep the FR coherence during the CZ crossing} 
\citep[][]{Fan09}. }

{In the solar atmosphere, some methods to estimate the twist rely on the computation of the helicity injection through the photosphere by shear motions and flux emergence  \citep{Demoulin2009,Liu2012,Thalmann2021}. 
However, it is not possible to recover the twist of a flux rope from direct photospheric observations, since the magnetic helicity is defined as a combination of field line topology and flux strength.
Recently, a different method has been proposed by \citet{MacTaggart21}, which consists of 
the magnetic helicity {without including the magnetic field strength.} 
This quantity, {defined as winding,} focuses on the field lines braiding 
and {it provides} a proxy to detect pre-twisted magnetic fields in emerging ARs.}

{Another important signature due to twist within these FRs} is the elongation of the magnetic polarities observed in LOS magnetograms during the AR emergence \citep[][]{Lopez-Fuentes00,Mandrini14,Dacie18,Lopez-Fuentes18}. 
This elongation is known as magnetic tongues and corresponds to the projection of the azimuthal component of the FR magnetic field in the LOS direction. 
Since the magnetic tongues affect the inclination of the polarity in version line (PIL), \citet{Luoni11} showed that the acute angle between the PIL and the bipole axis of an emerging bipolar AR determines the sign of the twist of the emerging FR. 
Moreover, \citet{Poisson15a} computed the acute angle between the PIL and the bipole axis measured from LOS magnetograms during the emergence of several ARs to estimate their twist. 
\citet{Poisson15b} compared these estimations with the twist obtained from coronal force-free field extrapolations, finding consistent results between both methods. 
These methods to estimate the twist are based on a FR model consisting of a half-torus magnetic field structure, in which the twist is determined by the number of turns of {the field lines around} the torus axis.

Photospheric observations of ARs have shown that they emerge slightly tilted, with their leading polarities closer to the solar equator than the trailing one. 
This tilt angle, and its spatio-temporal variation described by Joy's law \citep{Hale19}, are crucial for the Babcock-Leighton mechanism to model the global magnetic recycling process using current surface flux-transport models \citep[see][and references therein]{Cameron10,Yeates2023}. 
The tilt-angle dependence on the latitude has been confirmed by many authors
\citep[][]{Howard91,Wang91,Sivaraman93,Tian03} but with hemispheric and solar cycle variations \citep[see,\eg,][]{Tlatova18}. Variations in the results are also present depending on the method used to compute the tilt of bipolar ARs, {{\ie}, if it is done using} LOS magnetograms or white-light observations of sunspot groups \citep[see][and references therein]{Wang15}.

The longest data set for the latitudinal dependence of AR tilt is historically computed from white-light observations of sunspot groups \citep{Howard84}. 
The white-light tilt is estimated from the inclination {of the segment joining the centers of umbrae (and penumbrae) groups separated by polarity sign}. 
The information on the polarity sign {of each sunspot group is} obtained using different clustering methods or {directly} from magnetograms \citep[when they are available,][]{Baranyi16}. 
This method can produce wrong estimations of the tilt due to the construction of false bipoles, especially in the early phases of the AR emergence. 
An alternative estimation can be obtained from LOS magnetograms, despite the shortest time range of these observations (since the mid-1960s to the present). 
{In this case}, the magnetic tilt is computed from the magnetic flux-weighted center of the polarities, also known as magnetic barycenters. 
The tilt of the barycenters is defined as the acute angle between the segment that joints the AR magnetic barycenters and the equatorial plane. Differences between the white-light tilts and the {magnetic} barycenter tilts have been studied by \citet{Wang15} finding systematic deviations between both estimations.

Despite the robust estimation of the barycenter tilts, \citet{Poisson20a} have shown that this estimation can be strongly affected by the magnetic tongues during the emergence phase of ARs.
The position of the magnetic barycenters is shifted toward the PIL and the tilt is modified due to the asymmetric pattern of the magnetic tongues. 
When the tongues retract during the last half of the emergence, this tilt estimation produces a spurious rotation of the bipoles.
Similar effects were found for tilt estimations obtained with white-light data in \citet{Poisson2020b}. 
In {these} works we developed and tested a new method known as Core Field Fit Estimator (CoFFE) to compute the tilt angle of emerging ARs. 
The {CoFFE} method fits each magnetic polarity with a 2D Gaussian function, avoiding the flux close to the PIL, which is in general associated with the tongues. 
Results showed that CoFFE successfully reduced the effect of the tongues for the estimation of the tilt angle on eight different bipolar ARs.

Data modeling based on the Bayes theorem has expanded its applications into space science in the last decades due to the development of more efficient and generic computational tools \citep{Loredo92,Trotta08,Arellana23}. 
In solar physics, some previous applications of Bayesian inference were used to solve inverse problems associated with the propagation of MHD waves in coronal loops \citep[see][and references therein]{Arregui18}. 
In \citet{Poisson22}, we {used} a Bayesian method to model the LOS magnetograms of an emerging bipolar AR, and we {tested} the errors and stability of the method with 100 generated synthetic ARs. 
We found that the tilt angle inferred from these models successfully removed the effect of the magnetic tongues from the barycenter tilts, being even more robust than the estimations obtained with CoFFE.

In this work, we aim to obtain the magnetic parameters of the {FRs associated with four bipolar ARs, in particular their tilt angle and twist, with different inference schemes based on the evolution of LOS magnetograms}. 
These models include the temporal correspondence between the magnetograms, expanding the method proposed in \citet{Poisson22}.
{The global magnetic parameters obtained from these idealized models are useful to generate a reference for systematic comparison between statistical samples of ARs. In particular, the method presented here provides a novel estimation of parameters, such as the tilt angle, at the early phases of ARs lifetime.}
In \sect{data}, we summarize the standard tools used to process the LOS magnetograms. 
In \sect{ars}, we list and describe the observed properties of the selected ARs. 
\sect{model} contains a brief description of the half-torus FR model, and it presents the {methodology} 
of the proposed Bayesian temporal methods.
In \sect{params}, we present the results for the inferred parameters of the ARs using three different temporal methods and the method used in \citet{Poisson22}. 
In \sect{compare}, we compare the performance of the different methods with three different approaches: (1) analyzing the standard deviations between the model and the observations, (2) quantifying the correlation between the FR model parameters, and  (3) conducting a stability test over the methods when the information provided by the observations is synthetically reduced. 
In \sect{summary}, we summarize and discuss the main results obtained in previous sections, and in \sect{conclusions}, we conclude about the objectives and prospects.

 \section{Data Processing} \label{s:data}

%
 \begin{figure} 
 \centerline{\includegraphics[width=0.9\textwidth,clip=]{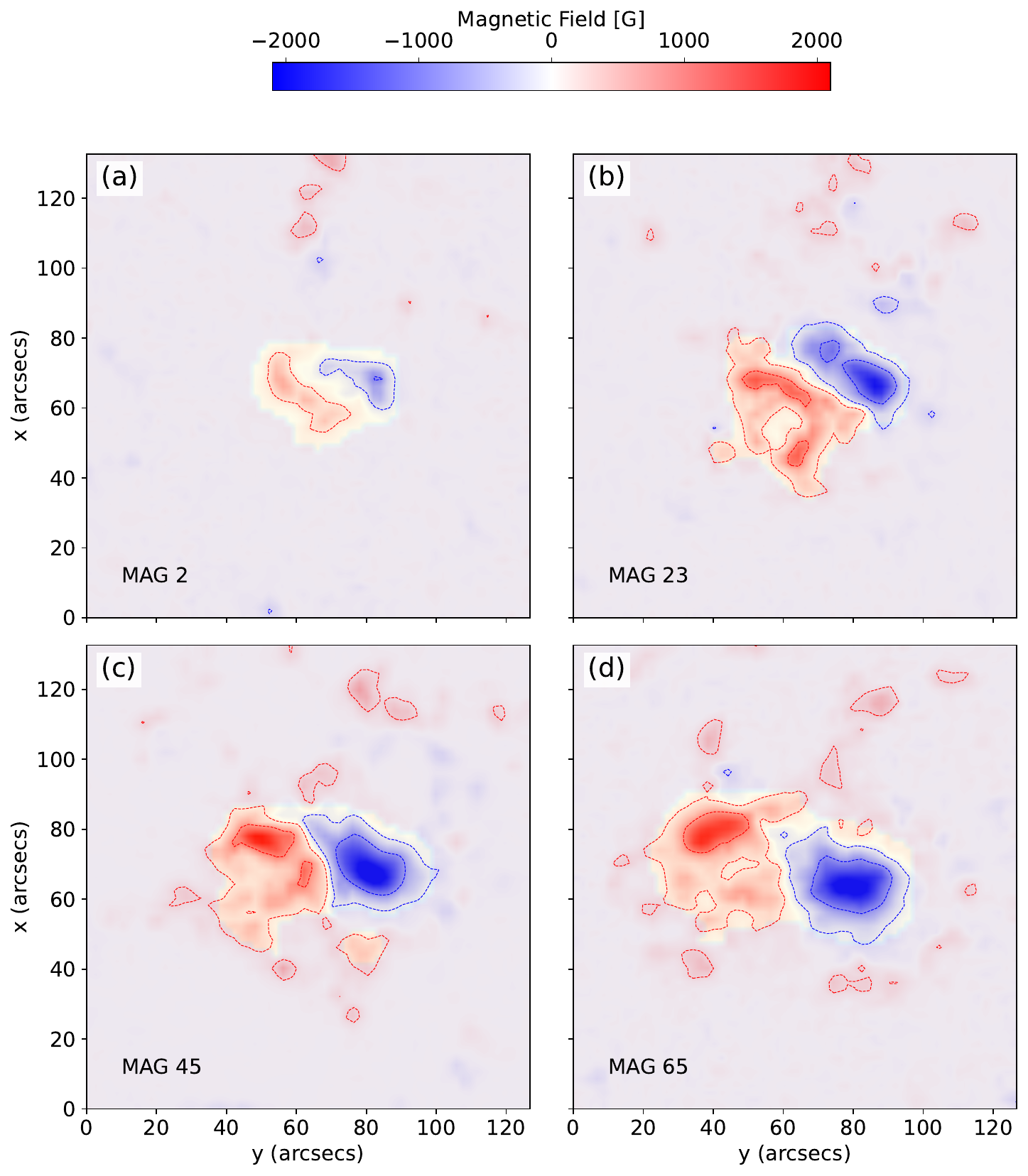}}
 \caption{SOHO/MDI LOS magnetograms for AR 10268. Each of the panels shows magnetograms for different times along its evolution. The \textit{red-} and \textit{blue-shaded} areas represent the positive and the negative LOS magnetic field components. The \textit{red} (\textit{blue}) contours correspond to the positive (negative) magnetic field with a strength of
200 and 1000 (-200 and -1000) G. The \textit{white} regions, including the inside magnetic polarities, show the mask used to reduce the effect of the background magnetic flux. These regions show the pixel selections used to apply the Bayesian method. A movie showing the evolution of this AR is available as
additional material (\href{run:./Movies/10268-mov.mp4}{10268-mov.mp4}).
} \label{fig:10268}
\end{figure}

We use line-of-sight (LOS) magnetograms from the 96-minute series obtained with the \textit{Michelson Doppler Imager} \citep[][]{Scherrer95} on board the \textit{Solar and Heliospheric Observatory} (SOHO). 
The photospheric magnetic field is mapped by measuring the Zeeman shift in the right and left circularly polarized light in a narrowband around $6768$ \AA~(Fe I line). The full-disk magnetograms from this series are constructed with 5-minute averaged magnetograms with a spatial resolution of $1.98''$ and an error per pixel of $\approx 9$ G \citep{Liu04}.

We perform the standard processing of LOS magnetograms using SolarSoftWare (SSW) tools.
For each of the analyzed ARs, we limit the observed time range to magnetograms in which the longitudinal position of the AR is within $-35^o$ to $35^o$ from the central meridian, to avoid projection effects of the LOS magnetic field \citep{Green03}. 
For each set of full-disk magnetograms, we transform the LOS component of the magnetic field to the solar radial direction by assuming that the photospheric magnetic field is radial. Then, we rotate the set of magnetograms to the time when the AR was located at the central meridian using the differential rotation coefficients derived by \citet{Howard90}. 
This allows us to select a sub-region that encompasses the AR evolution and to construct data cubes in time steps of 96 minutes (all available magnetograms within this MDI series).

To reduce the effect of the background magnetic field, we apply to the pixels a mask that evolves with the AR. The mask is constructed using tools from Python libraries Matplotlib and SciPy. 
We apply a uniform smoothing filter to the unsigned magnetograms and then select contours for a given field threshold. We visually inspect the masked field and select the threshold that best encloses the AR field. 
Once the mask is applied, we find a significant decrease in the imbalance between the positive and negative fluxes. This means that the flux imbalance is mainly due to the background field present before the AR emergence. 
To minimize this background contribution, every pixel outside the mask is set to zero. Figure~\ref{fig:10268} shows an example of AR 10268 masked region at different times. Gray-shaded regions indicate the mask complement and therefore the pixels that are set to zero. 

\section{Selected ARs} \label{s:ars}

In this section, we briefly describe the observed magnetic field properties of the four studied ARs. These ARs are selected as representative cases of isolated bipolar regions in which their evolution is observed from the first flux emergence until the maximum flux is reached and the decaying phase starts. This time range and the corresponding AR heliocentric coordinates fulfill the longitudinal and latitudinal restrictions mentioned in the previous section. For all these ARs the cadence between each magnetogram is 96 minutes without any gap or missing data along their full analyzed evolution. Therefore, we will use the number of magnetograms for each AR, instead of dates or time steps, to describe their evolution. 

%
 \begin{figure} 
 \centerline{\includegraphics[width=0.9\textwidth,clip=]{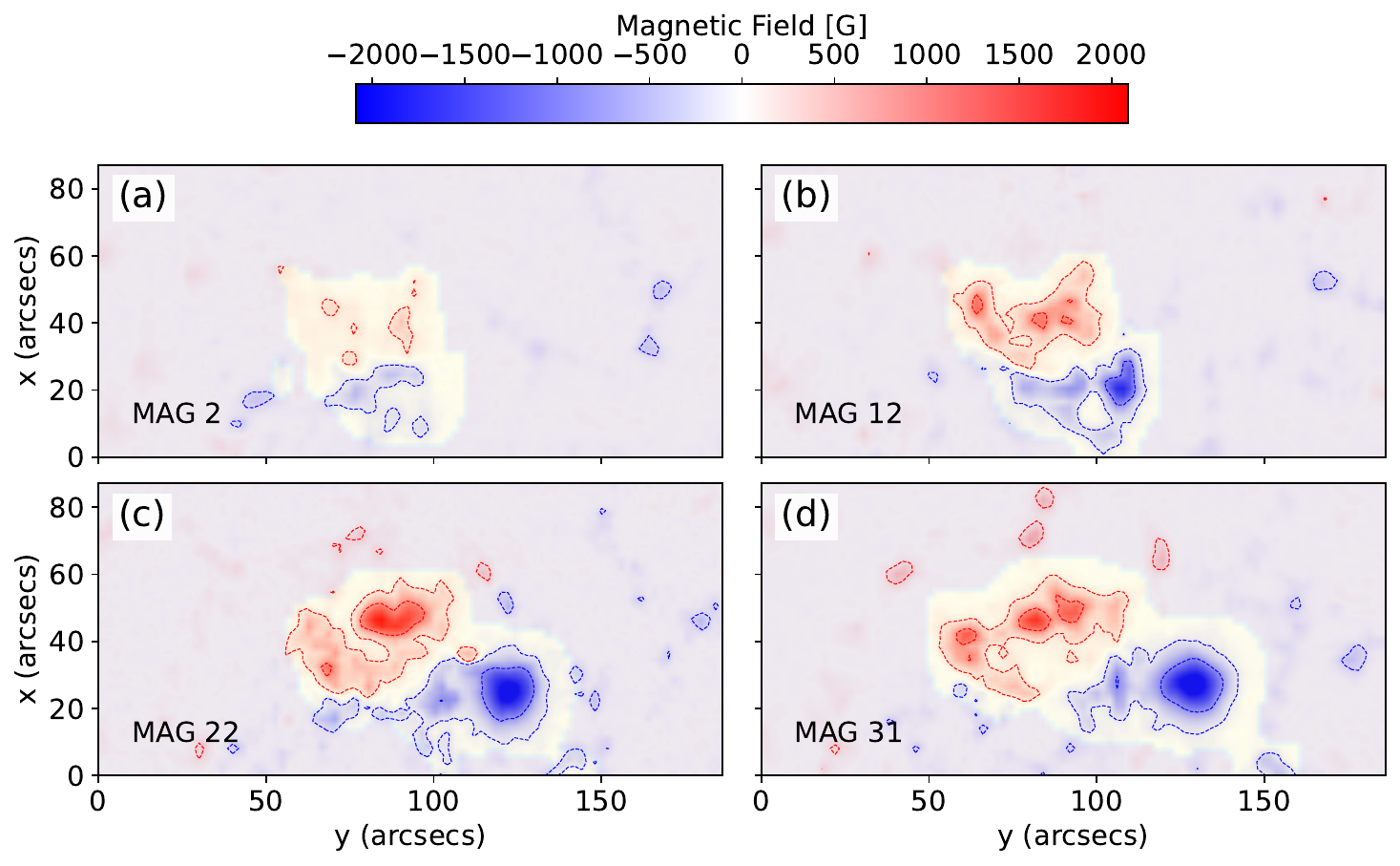}}
 \caption{SOHO/MDI LOS magnetograms showing the evolution of AR 10274 with the same color convention used in \fig{10268}. A movie showing the evolution of this AR is available as
additional material (\href{run:./Movies/10274-mov.mp4}{10274-mov.mp4}).
} \label{fig:10274}
\end{figure}

AR 10268, previously studied by \citet{Poisson22}, is a $\beta-$type AR located in the northern hemisphere (N16). It presents a five-day-long emergence starting on January 21, 2003. 
Figure~\ref{fig:10268} shows four panels at different stages of the AR evolution.
The leading negative polarity conserves a good cohesive shape with clear elongated polarities throughout the evolution. The elongation of the polarities on this AR corresponds to a clear example of the definition provided in \sect{intro} for the magnetic tongues. In this AR, the positive following polarity is more dispersed and has a more fragmented tongue than the leading one. Both polarities are well distinguished from the background positive magnetic flux. In \citet{Poisson2020b}, we found that the estimation of the tilt angle of this AR is strongly affected by the magnetic tongues, and corrections obtained using CoFFE present differences of more than 20$^\circ$ with a direct barycenter estimation. This AR is a good example for comparing with previously developed methods to estimate parameters such as the tilt angle and the number of turns. The full evolution of the AR can be found in the movie called 10268-mov.mp4 in the supplementary material.

%
 \begin{figure} 
 \centerline{\includegraphics[width=0.9\textwidth,clip=]{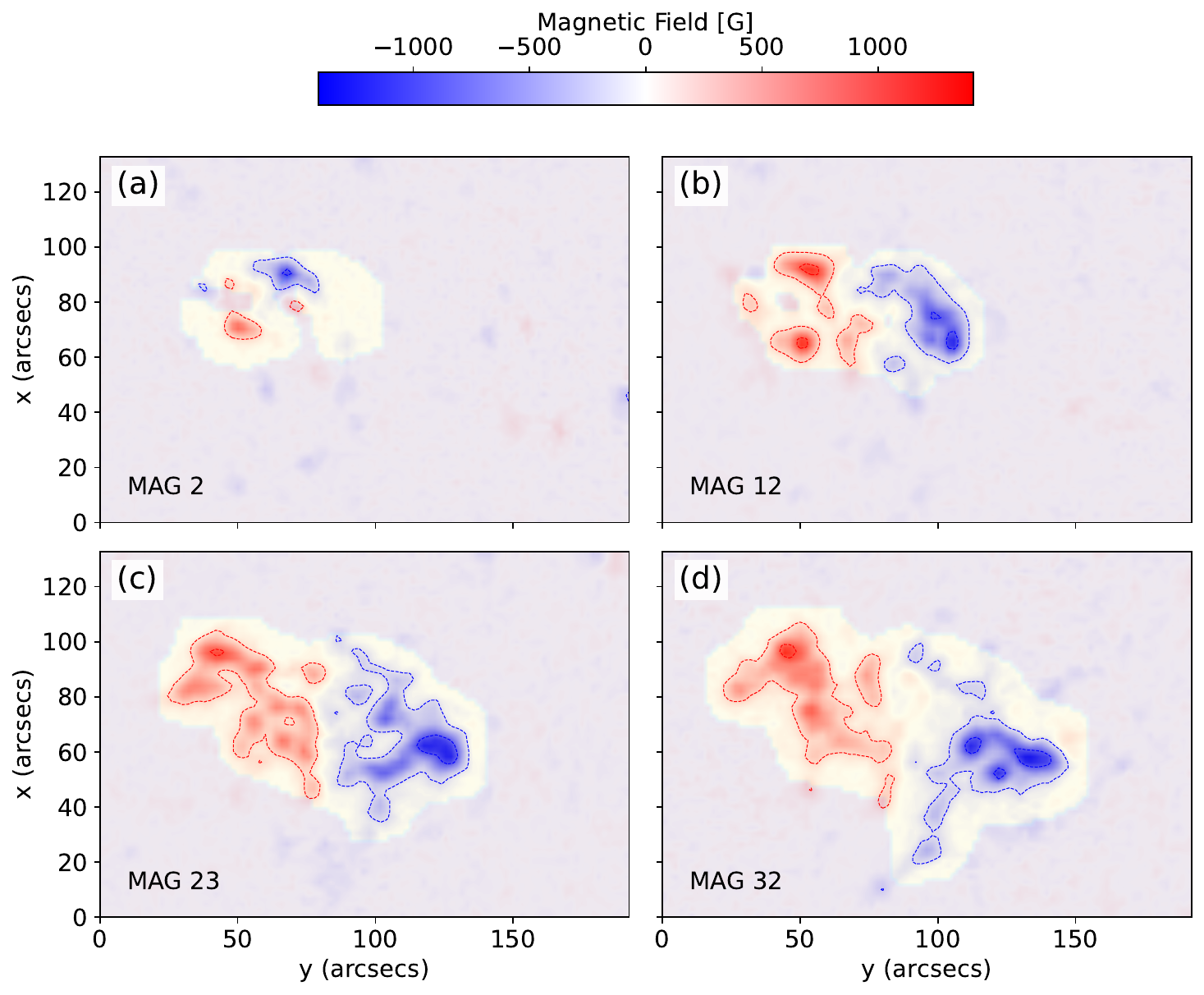}}
 \caption{SOHO/MDI LOS magnetograms showing the evolution of AR 8056 with the same color convention used in \fig{10268}. A movie of the evolution of this AR is available as
additional material (\href{run:./Movies/8056-mov.mp4}{8056-mov.mp4}).
} \label{fig:8056}
\end{figure}

AR 10274 appeared in the southern hemisphere (S07) on February 1, 2003. We follow the emergence of this bipolar AR up to February 3 (counting 33 magnetograms) when dispersion starts to have a clear effect on the flux distribution. 
Different stages of the emergence can be seen in the panels of Figure~\ref{fig:10274}.
It is a small AR with long but fragmented magnetic tongues in both polarities. The computed mask allows us to isolate the core flux of both polarities from the background positive flux. 
This AR is highly tilted concerning the east-west direction, having an opposite inclination to Joy's law. This example will be used to test our model for a case in which dispersion and fragmentation of polarities take place during the AR emergence. The AR fully analyzed evolution can be seen in the supplementary movie called 10274-mov.mp4.

AR 8056 is a low-flux bipolar AR that emerged in the northern hemisphere (N17) in June 1997. Due to the latitudinal constraints described in the previous section, we analyze only a part of the AR evolution from mid-June 24 up to late June 26, counting 32 magnetograms. 
Panels in Figure~\ref{fig:8056} show examples of magnetograms corresponding to the AR evolution.
The AR maximum flux is reached at magnetogram 30. At the last part of the emergence, fragmentation of the positive following polarity becomes stronger, suggesting the start of the decaying phase of the AR flux evolution.
Tongues are weak and detected only during a short span of the full emergence. The pattern displayed by the tongues corresponds to the emergence of a negatively twisted FR using the criteria presented in \citet{Luoni11}, where the twist sign is derived from the acute angle formed by the PIL and the bipole axis. During the first half of the emergence, this AR presents the strongest clockwise rotation of the polarities, which is not associated with the elongation or contraction of the magnetic tongues.  The AR full evolution can be seen in the supplementary movie called 8056-mov.mp4.

%
 \begin{figure} 
 \centerline{\includegraphics[width=0.9\textwidth,clip=]{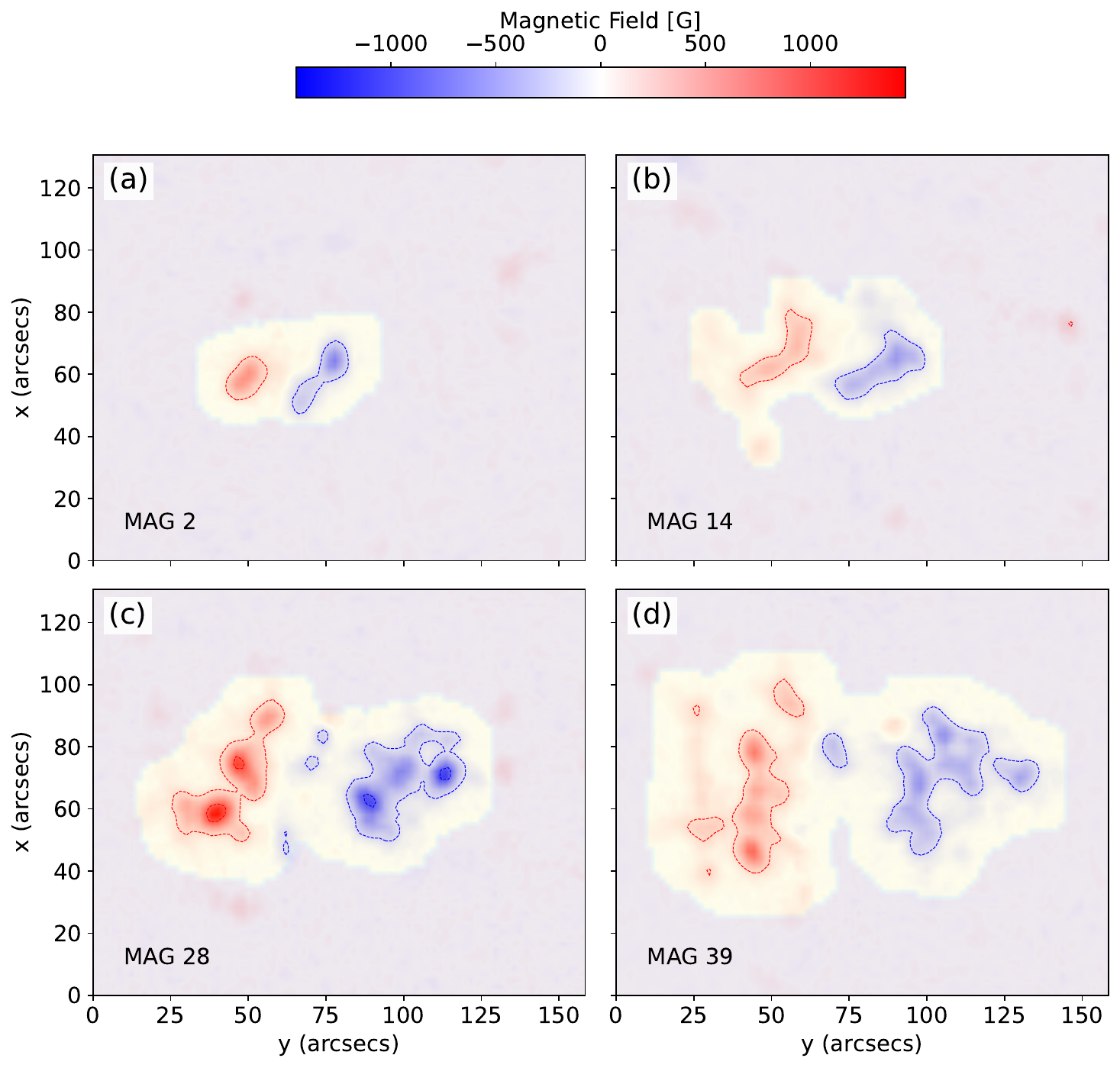}}
 \caption{SOHO/MDI LOS magnetograms showing the evolution of AR 8060 with the same color convention used in \fig{10268}. A movie showing the evolution of this AR is available as
additional material (\href{run:./Movies/8060-mov.mp4}{8060-mov.mp4}).
} \label{fig:8060}
\end{figure}

AR 8060 emerged with a simple bipolar flux configuration in the northern hemisphere (N05) close to the solar equator. We analyze 41 magnetograms observed between July 7 and 10, 1997. 
\fig{8060} shows four panels with different magnetograms of AR 8060.
The AR maximum flux is reached around magnetogram 32, and we extend the analyzed time range until decaying flux effects become noticeable. Tongues produce a mild elongation of the polarities, with a pattern well in agreement with a positive twisted FR. The AR polarities conserve a cohesive shape during the evolution, with no significant fragmentation.  The fully analyzed evolution can be seen in the supplementary movie called 8060-mov.mp4.

\section{Model {and Methods Descriptions}} \label{s:mometh} 

The model for generating synthetic magnetograms is based on the emergence of a sub-photospheric half-torus FR with uniform twist. This model considers only the magnetic field geometry, and it does not include the deformations and reconnections occurring during emergence or any interaction between the magnetic field and the plasma. Despite this simplification of an AR emergence, the model is aimed to reproduce global aspects of the photospheric magnetic flux distribution of $\beta$-type ARs. In this work, we limit our analysis to this particular type of ARs, which indeed correspond to the most commonly observed ones during any solar cycle stage.

\subsection{{Magnetic Field Model}} \label{s:model} 

The FR is a 3D half-torus magnetic field structure defined by four independent parameters, here called ``field" parameters: a small radius $a$, a large radius $R$, an axial field strength $B_0$, and a twist parameter $N_t$. $a$ and $R$ are associated with the size and separation of the magnetic polarities, respectively, so they have spatial units (such as pixels or arc seconds). $B_0$ indicates the field strength at the FR central axis in Gauss units. The model uses a symmetric axial field distribution with a Gaussian profile centered at the axis, in which $B_0$ is the maximum field strength and the flux spreads around its center with a standard deviation $\sigma = a$. The total axial flux can be written as $\Phi_A = \pi B_0 a^2$. Due to its importance and physical significance, we will use $\Phi_A$ as our FR model parameter instead of $B_0$, which is more difficult to link to any observed property of AR magnetograms. Finally, the twist parameter $N_t$ corresponds to the number of turns of magnetic field lines around half of the torus axis. This non-dimensional parameter provides the signed amount of the FR twist \citep[see the appendix in][for details of this model]{Poisson22}.

Projecting the FR field onto the normal to successive horizontal planes representing the photosphere located at different heights in the direction of emergence \citep[see][]{Poisson22}, the model produces synthetic magnetograms that emulate the flux emergence of an AR. The relative position between these photospheric planes and the FR includes four positional parameters of our model: the half-torus center position at a depth $d$, its tilt $\alpha$, and {the horizontal Cartesian coordinates of the FR center} $x_c$ and $y_c$. $d$ indicates the vertical position of the plane, so it gives a measure of the stage of the AR emergence. 
{We define the dimensionless $d_0$ as}
  \begin{equation} \label{eq_dnot}
  d_0 = 1-d/(R+a){\rm,}     
  \end{equation}
\noindent which indicates the fraction of the FR that is above the photospheric plane, scaled with the distance between the torus center and the FR apex. Therefore, the emergence of the FR starts with $d_0\approx 0$ and ends with $d_0 = 1$. 

Each generated synthetic magnetogram is defined by a vector with the values of the above-defined parameters $\vp = (a,R,N_t,\Phi_A,d_0,\alpha,x_c,y_c)$. To model a set of $N$ magnetograms corresponding to the emergence of an AR, we can consider independence between the parameters obtained from different magnetograms. 
In this case, we will generate a sequence of magnetograms corresponding to $N$ different sets of parameters. This approach is equivalent to the method used by \citet{Poisson22} since the method was designed to model individual magnetograms. In \sect{method}, we describe alternative methods in which a temporal correlation is introduced, reducing the dimension of the parameter space.

\subsection{Inference Methods} \label{s:inference}

In the following analysis, there are two sets of statistical variables: first the FR parameters $\vp$ and second the magnetic field values of the observed magnetogram $\Mo$.
Both have probability distributions that are related to the following process of linking the FR model to the observations. 
Our method is based on the Bayes theorem \citep{Bayes}:
  \BE \label{eq_Post}
   P(\vp|\Mo)  =  \frac{ P(\Mo|\vp) \, P(\vp)}{P(\Mo)}  \,,
  \EE
\noindent $P(\vp|\Mo)$ is the conditional probability distribution of the FR parameters, $\vp$, for a given set of observed magnetograms, $\Mo$. $P(\Mo| \vp)$ is the conditional probability to get the observed magnetogram, $\Mo$, for a set of parameters $\vp$. $P(\Mo)$ is the probability distribution of observing $\Mo$ (with all the possible $\vp$ parameters), while $P(\vp)$ is the probability distribution of a set of  $\vp$ parameters (for all possible $\Mo$).
{\eq{Post} expresses that the probability to have both specific values of $\Mo$ and $\vp$ could be computed in two equivalent ways: by selecting $\Mo$ with the probability $P(\Mo)$ and multiplying it by the conditional probability $P(\vp|\Mo)$, or by selecting $\vp$ with the probability $P(\vp)$ and multiplying it by the complementary conditional probability $P(\Mo|\vp)$).}

For a set of observed magnetograms, $\Mo$, we want to derive the probability distribution of the parameters $\vp$ of a FR model which could represent this $\Mo$.  This corresponds to derive the posterior distribution, $P(\vp|\Mo)$.  
\eq{Post} indeed relates $P(\vp|\Mo)$ to the prior distribution, $P(\vp)$, which contains our knowledge on the parameters $\vp$. 
The marginal likelihood $P(\Mo)$ in \eq{Post}, describes the probability distribution of observing $\Mo$ with all the possible parameters, $\vp$, as defined based on the prior probabilities. For a given {set of observed magnetograms} $\Mo$, this distribution is the same for all combinations of $\vp$, then it does not affect the relative probabilities of different parameters, and it acts as a normalizing factor for the posterior $P(\vp|\Mo)$. 
Since $P(\vp|\Mo)$ is an eight-dimensional distribution, to simplify our analysis, we can compute a ``marginal posterior" for each parameter.  These distributions (one for each FR model parameter) are obtained by integrating the full posterior in \eq{Post} over the remaining seven parameters within the prior bounds. Below, we simply call it the parameter posterior, and for a parameter $p$, we write it $P_m(p|\Mo)$.

The conditional probability $P(\Mo|\vp)$ is set to the likelihood function $\cL$:
  \BE \label{eq_likelihood}
  \cL(\Mo;\Mpi,\sigma) = \frac{1}{\sigma\sqrt{2\pi}} 
            \exp \Bigg(-\frac{1}{2 \sigma^2} \sum_i (\Mo-\Mpi)^2 \Bigg) \,,
  \EE

\noindent in which we assume the errors, between the modeled data cube $\Mpi$ and the observed one $\Mo$, are characterized by a normal distribution with a zero mean and a standard deviation $\sigma$. This hypothesis is based on the assumption that the instrument is well calibrated, without any bias or systematic errors, and that the model matches perfectly the observations up to the presence of random errors (in particular we do not account for systematic bias). We assume that the errors between the model and the observations are only due to a random process that is independent for each magnetogram pixels, so the probabilities of individual pixels are multiplied to compute the magnetogram probability.  This implies the summation over the pixels present in the exponential of \eq{likelihood}.
With this hypothesis, the likelihood function of \eq{likelihood} expresses the probability of getting the magnetogram $\Mo$ with the given parameters $\vp$. Therefore, $\sigma$ corresponds to the mean error between the model and the observations, {\eg}~ if the model perfectly describes the observations $\sigma$ will be the instrumental error ($9$ G for this MDI series). The value of $\sigma$ will impact directly on the width of the posterior $P(\vp|\Mo)$. If $\sigma$ is small enough, \eq{likelihood} defines a function with narrower peak(s) in $\vp$ than $P(\vp)$. This is the desired condition for this study.  In such case, $\cL(\Mo,\Mp,\sigma)$ is the main contribution to the posterior $P(\vp,\Mo)$ that we are searching for.

The $P(\vp)$ distribution contains all the a priori physical information known about the FR model parameters. The priors guide the sampling process into physically reasonable ranges. For our model, we choose uniform distributions for all parameters except for $\Nt$ (as explained below). The uniform distribution sets boundaries to the parameter space without imposing any particular weight. Despite
the simplicity of the uniform prior, we expect that, if the model provides a good approximation to the observations, the inference process should converge to a proper posterior distribution independently of the selected prior. We set the prior ranges automatically from direct measurements made over the data cubes. 

For $a$, we compute the mean size of the polarities, as done in \citet{Lopez-Fuentes00}. Then we take the maximum and minimum value of these sizes during the AR emergence, and we multiply the upper and lower boundary by a factor of $1.5$ and $0.5$, respectively. This produces {an extended} 
prior for the parameter $a$, assuring the existence of a maximum likelihood region within the defined interval.
We perform a similar range selection for $R$, using the distance between the magnetic barycenters of each polarity, and for $\Phi_A$, using the unsigned magnetic flux. Tilt angle ranges are defined between $-\pi/2$ and $\pi/2$, and $d_0$ between $0$ and $1$. 

We select a narrow range of $6$ by $6$ pixels for the FR center, $x_c$ and $y_c$, around the position of the AR unsigned magnetic flux barycenter. By limiting the ranges for these parameters, we significantly improve the method performance, reducing the size of the sampled parameter space. Our model does not take into account the magnetic flux imbalance or asymmetry between the magnetic polarities, so the FR central position is expected to be located at the flux-weighted center. Despite this condition may not be fulfilled by real ARs, the inferred parameters $x_c$ and $y_c$ are expected to be well approximated by the AR magnetic flux center. In \citet{Poisson22}, we found that this approximation was good within an error of $\approx 2$ pixel, {so this is our choice of narrow priors for $x_c$ and $y_c$.}

We define the prior of $\Nt$ by combining the information of the sign of the twist and its magnitude. For the twist sign, we use a binary distribution in which both possible values, $1$ or $-1$, have equal probability ($0.5$ for each). Since the prior for $|\Nt|$ should be strictly positive, we select a gamma distribution with shape and scale parameters of $3$ and $0.1$, for which the distribution mean is $0.3$ and a standard deviation is $\approx0.2$. This distribution is skewed toward small values and, without an upper boundary. The probability quickly tends to zero for large values. The use of this distribution is based on results presented in \citet{Poisson15a} in which $\Nt$ values were estimated from the PIL inclination of 41 ARs. 

We use Python 3.6 library PYMC 5 for sampling the posterior in \eq{Post} \citep{PyMC3}. The inference procedure requires the evaluation of the likelihood function in \eq{likelihood} for different sets of parameters. The computation is optimized by generating tensor graphs for this function, which is done with the library Pytensor, which efficiently evaluates mathematical expressions involving multidimensional arrays. 
Then, we need to select a proper sampler to explore the parameter space efficiently. We use the No-U-Turn sampler (NUTS). This algorithm is a particular Hamiltonian Monte Carlo method, in which a random walk behavior is avoided by stopping automatically the sampling when it starts to double back and retrace its steps \citep{NUTS}. 

We use the same standard setup for NUTS in all methods, consisting of four independent chains with 2000 samples each. Each chain starts at different points within the parameter space, so there is no correlation imposed between chains. Convergence to a single posterior takes place when the four chains sample the same distribution. We test convergence using the normalized rank diagnostic tests, $\hat{R}$ \citep{Vehtari2021}. This rank is simply the ratio of the mean of the variance for each chain (2000 samples) and the variance between the combined chains (8000 samples). 
If convergence is achieved, then these variances (the within-chain and between-chains) should be identical for each chain, so $\hat{R} = 1$. If chains have not converged to similar distributions, then different distributions are obtained with all or some chains, and the values of $\hat{R}$ for each chain increase above unity. This analysis establishes a limit of 1.4 to consider a reasonable convergence. Above this value, the different chains converge to different distributions. For all the models evaluated in this work, we found values of $\hat{R}$ below $ 1.1$ indicating a convergence of NUTS to a single posterior.

\subsection{Methods Based on the Definition of Priors} \label{s:method} 

In \citet{Poisson22}, we inferred the FR model parameters for all the LOS magnetograms of AR 10268 emergence. The developed method can be used to model individual magnetograms. We will refer to this method as the Temporal Method -0 (TM-0): a temporal method with no imposed constraints on the parameters. 
The TM-0 fits the observations with the maximum number of free parameters ($8\times N$ parameters) neglecting any temporal correlation or coherent evolution of the parameters of the model.

As an alternative approach, we can produce Temporal Methods (TMs) in which some of the parameters remain constant along the temporal dimension. For instance, setting as constant all four field parameters described in \sect{model} ($a$, $R$, $\Nt$, and $\Phi_A$), but allowing the evolution of the four positional parameters ($d_0$, $\alpha$, $x_c$, $y_c$, ),  we generate a method that describes the AR emergence with a kinetic translation and rotation of the FR as a whole (keeping its magnetic structure unchanged). We will refer to this method as the TM-4 (the number refers to the parameters held constant). The total number of parameters is significantly reduced to $4\times N + 4$, being the minimum number of parameters considered in this study.

In this sense, we can create multiple TMs between TM-4 and TM-0, by defining any of the field parameters as a {unique scalar or as a} $N-$ dimensional vector. {For simplicity, we will focus only on two additional methods, one in which $a$ can evolve, called TM-3 (3 because $R$, $\Nt$, and $\Phi_A$ are set as constant during the emergence), and the other in which $\Nt$ and $\Phi_A$ are set as constant during the emergence, called TM-2. We choose those two field parameters, $a$ and $R$, to be capable of evolving since there is a priori no indication that they have to be constant during the AR emergence. Moreover, it is expected that during the emergence the evolution of $a$ and $R$ represent possible expansions of the FR cross-section and the FR length once it emerges above the photosphere due to changes in the surrounding physical conditions. On the other hand, $\Nt$ and $\Phi_A$ are selected as constants for most TMs since they are conserved quantities in ideal MHD conditions (no reconnection). For a summary of the methods and their defined parameters see Table \ref{table0}.} 

  \begin{table} 
\caption{List of constant and variable parameters along the temporal dimension for each method.
} \label{table0}
\begin{tabular}{ccc} 
  \hline
Method & Constant & Variable  \\
  \hline
TM-0 &   & $a$, $R$, $\Nt$, $\Phi_A$, $d_0$, $\alpha$, $x_c$, $y_c$ \\
TM-2 & $\Nt$, $\Phi_A$  & $a$, $R$, $d_0$, $\alpha$, $x_c$, $y_c$  \\
TM-3 & $R$, $\Nt$, $\Phi_A$  & $a$,  $d_0$, $\alpha$, $x_c$, $y_c$  \\
TM-4 & $a$,$R$, $\Nt$, $\Phi_A$  &   $d_0$, $\alpha$, $x_c$, $y_c$  \\

  \hline
\end{tabular}
\end{table}


As mentioned in the previous section, $\sigma$, the standard deviation of the likelihood function of \eq{likelihood}, has a direct impact on the broadness of the posterior, but it is expected that the most probable value of the distribution will not be affected if convergence to a single point in the parameter space is achieved. Still, a proper selection of $\sigma$ will give us the precision at which the different parameters of the model are estimated. Since our model is meant to describe only the global aspects of the magnetic field distribution, without considering deformations and interaction with the surrounding plasma, it is reasonable to use a value of $\sigma$ larger than the instrumental error of $9$ G. Indeed, the mid- and small-scale features are not represented with the half-torus model. In \citet{Poisson22}, we estimated $\sigma$ as the mean difference between the magnetograms and the best model (usually obtained with a different $\sigma$ value). This estimation results in a $\sigma$ that scales with the total magnetic flux, ranging from $\approx30$ G up to $\approx150$ G for AR 10268. 

In this work, we will use a different estimation, giving a single value of $\sigma$ for each data cube. For that, we look at the mean background {magnetic field} 
since this random network field is not included in our current model, and it is a measurable quantity for each case. To estimate the background magnetic field, we use the mask complement field, and we compute its standard deviation around zero. This value gives us information about the surrounding mean field conditions. The selection of this value for $\sigma$ relies on the approximation that the field of the AR polarities will be affected by the background network similarly to its surroundings. This approach provides a measurable $\sigma$ that could be underestimated. Still, for the aim of the present work, it will be enough, since we are not interested in the magnitude of the parameters but in comparing the different TMs using common measurable standard deviations. In all cases, we found that $\sigma$ ranges between $40$ G up to $60$ G which is around the standard deviation of the difference between the best models and the observed magnetograms. This further justifies our way of defining $\sigma$.

 \begin{figure} 
 \centerline{\includegraphics[width=0.95\textwidth,clip=]{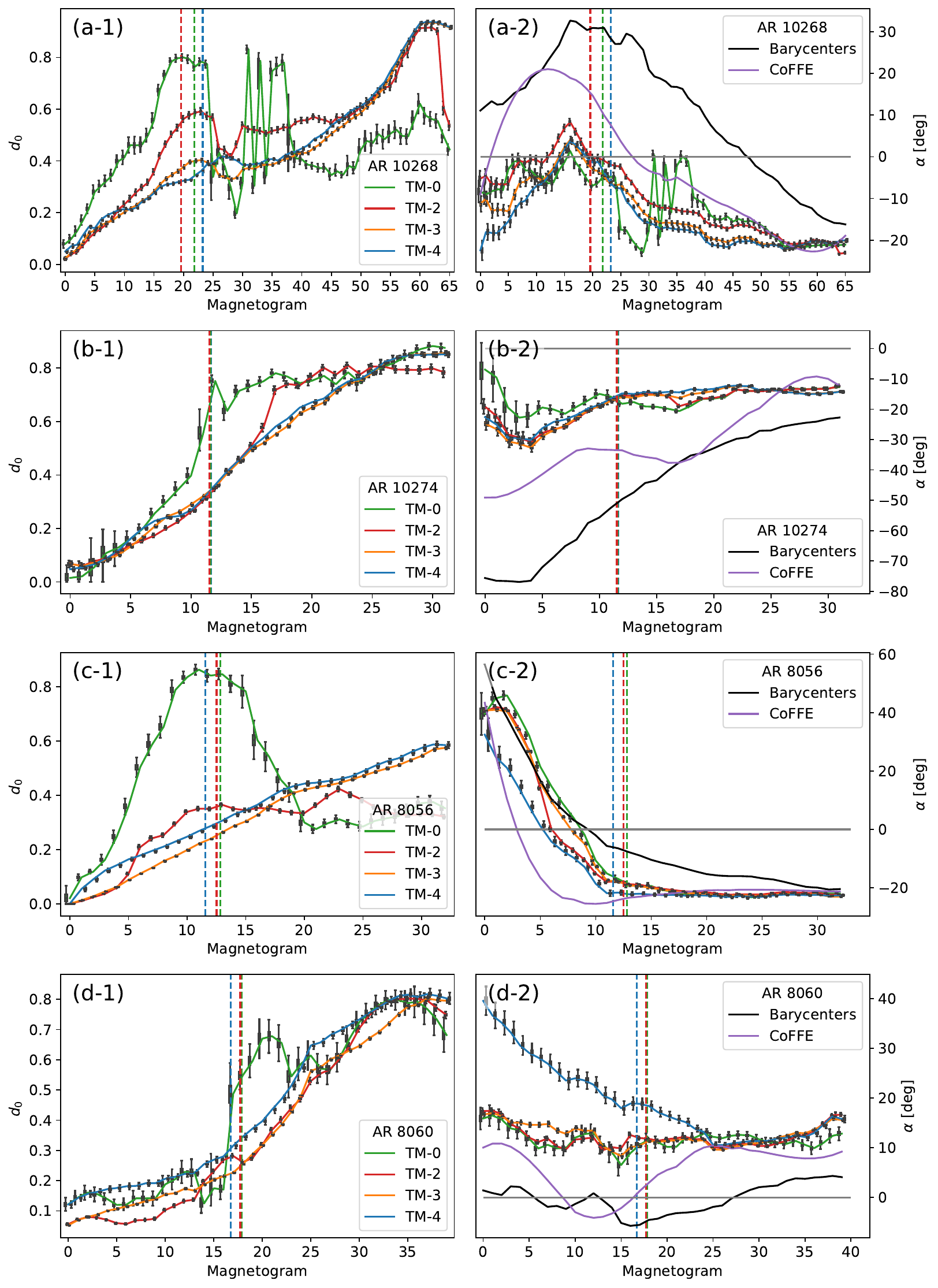}}
 \caption{Evolution of the inferred posteriors for the field positional parameters $d_0$ (\textit{left column}) and $\alpha$ (\textit{right column}) for ARs 10268 \textbf{(a)}, 10274 \textbf{(b)}, 8056 \textbf{(c)}, and 8060 \textbf{(d)}. \textit{Boxes} at each time correspond to the quartile of the marginal distributions obtained with TM-0 (\textit{green}), TM-2 (\textit{red}), TM-3 (\textit{orange}), and TM-4 (\textit{blue}). \textit{Whiskers} indicate the extension of these posteriors {up to $1.5$ of the interquartile range}. The corresponding colored lines mark the evolution of the median of these distributions, respectively. \textit{Vertical dashed lines} indicate the interpolated time $t_c$ in which the depth of the FR torus center, $d$, is equal to $R$ for the TMs. \textit{Black} and \textit{violet lines} in the tilt plots indicate the tilt estimations using the magnetic barycenters and the COre Field Fit Estimator (CoFFE) methods, respectively.
} \label{fig:AR-params1}
\end{figure}

\section{Inferring the Parameters of the Model} \label{s:params} 

\subsection{Positional Parameters} \label{s:pparams}

 The FR location and orientation are defined by four parameters: $d_0$, $\alpha$, $x_c$, and $y_c$. Since $x_c$ and $y_c$ are well defined (up to $\pm 1$ pixel), we summarize below the results obtained with the normalized depth $d_0$ and the tilt $\alpha$. 
 
 \Fig{AR-params1} shows the parameter posterior  of the positional parameters, $P_m(d_0|\Mo)$ (left column) and $P_m(\alpha|\Mo)$ (right column), for all analyzed ARs. Each row corresponds to a different AR, being panels a to d the results obtained for ARs 10268, 10274, 8056, and 8060, respectively. On each panel, the boxes show the evolution of the quartile of the posterior, and the whiskers mark the extension of the distribution up to $1.5$ of the interquartile range. Colors indicate the distributions obtained for the TM-0 (green), the TM-2 (red), the TM-3 (orange), and the TM-4 (blue). The colored solid lines mark the evolution of the median for each model. 

The evolution of the median for $d_0$ is significantly different between the TM-0 and all the other TMs, being the monotonic increase of $d_0$ the expected evolution for a kinetic rise of the FR as a whole. The fluctuation on $d_0$ for the TM-0 reflects the lack of coherence for the FR vertical position at different times. Despite its limitations, TM-0 estimates a value of $d_0$ which is consistent with the other TMs at the early phase and the last phase of the emergence of ARs 10274 (panel b-1) and 8060 (panel d-1). The other three TMs (TM-2, TM-3, and TM-4) present a similar monotonic evolution of $d_0$, being TM-2 the most fluctuating for ARs 10268 (panel a-1) and 8056 (panel c-1). This difference can be due to an important correlation between $d_0$ and $R$ in TM-2 in contrast to TM-3 and TM-4 (this will be analyzed in \sect{corr}).

The dashed vertical lines correspond to the times $t_c$ when the axis top crosses the photosphere, $\ie$ the FR cross-section at its top is half above and half below the photosphere. Then, around this time the synthetic magnetogram is the most sensitive to changes in $d_0$. Indeed, we find some changes in the emergence rate seen on the slope of $d_0$, which coincides with these times. The changes are stronger when the FR model parameters are less constrained, especially large for TM-0. The most significant changes in the emergence rate are seen for ARs 10268 (panel a-1) and 8060 (panel d-1). 

For all cases, in \fig{AR-params1}, we show the period corresponding to the AR emergence defined as follows. We look at the observed LOS magnetic flux of the AR until it reaches its maximum. We include several magnetograms after the maximum flux as shown in \fig{AR-flux} (when the longitudinal range criterion defined in \sect{data} allows us to do it) because we have found using the FR models that the azimuthal flux of the tongues can contribute to the AR observed total flux, producing a peak in the flux before the end of the emergence \citep[see Figure 9 in][]{Poisson16}. We follow the evolution of the AR after the maximum flux peak until the effects of decay and fragmentation of the polarities become noticeable. Despite this selection, we find that all the $d_0$ obtained with the TMs reach maximum values below unity. For ARs 10268 (panel a-1), 10274 (panel b-1), and 8060 (panel d-1) the emergence ends when $d_0$ is just above $0.8$. AR 8056 reaches values of $d_0$ around $0.6$ when decaying effects impact the main polarities. This saturation limit of $d_0$ can be associated with the physical process in which the top of the FR is above the photosphere, and hence, the FR stops being buoyant.

The right column of \fig{AR-params1} shows the evolution of the tilt posteriors for the corresponding TMs and ARs. These plots include estimations of the tilt obtained with the magnetic barycenters and the COre Field Fit Estimator \citep[COFFE;][]{Poisson20a} plotted with black and violet lines, respectively. For ARs 10268 (panel a-2), 10274 (panel b-2), and 8056 (panel c-2) the four TMs (TM-0 to TM-4) provide similar estimations of the tilt angle evolution. In the case of AR 8060 (panel d-2), TM-4 has the largest variation of all TMs, suggesting a large counter-clockwise rotation during the first half of the emergence. Later on, all TMs converge to the same tilt values. The tilt becomes more stable after the time $t_c$ (vertical dashed lines) in all cases. 

{The different estimations of the tilt converge towards the end of the emergence, indicating that the main effect of the tilt dispersion {between different methods} is associated with the elongation of the magnetic tongues as seen in previous works \citep{Poisson2020b}. ARs 10268 and 10274 present strong elongated tongues during most of their evolution shifting the magnetic barycenters towards the PIL. This effect produces a spurious rotation when the tongue retracts and loses intensity toward the end of the emergence. In consequence, the tilt obtained using the barycenters is strongly affected by the tongues (see the black line in the right panels of \fig{AR-params1}). The CoFFE method (violet line) partially removes the effect of the tongues.  It does so by selectively focusing on the most symmetrical aspects of the polarities at each given time while ignoring the fields closer to the PIL. CoFFE provides an intermediate solution between the TMs and barycenters tilt.} 

{The TMs reproduce the global elongation of the polarities, so the tilt angle obtained is due only to the intrinsic inclination of the inferred FR. Corrections on the tilt obtained with the TMs are significantly large for ARs 10268, 10274, and 8060, where the tongues are more prominent. On the other hand, the less elongated polarities observed for AR 8056 correspond consistently well with the tilt over all methods.} 

The tilt evolution obtained with TM-4 for AR 8060 presents a significant difference among all other estimations (see \fig{AR-params1}d-2). 
This difference can be explained because of the inferred parameter $\Nt$ for TM-4 is $50\%$ larger than the estimations given by TM-3 and TM-2 (see Table~\ref{table2}). This larger twist is used to model the 
elongated polarities observed at the early phase of the AR emergence (see \fig{8060}).
The TM-4 generates an intrinsic rotation of the tilt to compensate for the contraction {with time} of the inferred strong magnetic tongues. 
In the case of the TM-3 and the TM-2, the evolution of the polarities at the beginning of the AR emergence is modeled with an increase of the parameter $a$ (not shown here), instead of a large $\Nt$ as for the TM-4 (where $a$ is imposed to be constant). Therefore, the tongues in these models do not produce any significant spurious rotation of the bipole. We conclude that imposing a constant radius could lead to biased results.

\subsection{Magnetic Field Parameters}

 \begin{figure} 
\centerline{\includegraphics[width=0.95\textwidth,clip=]{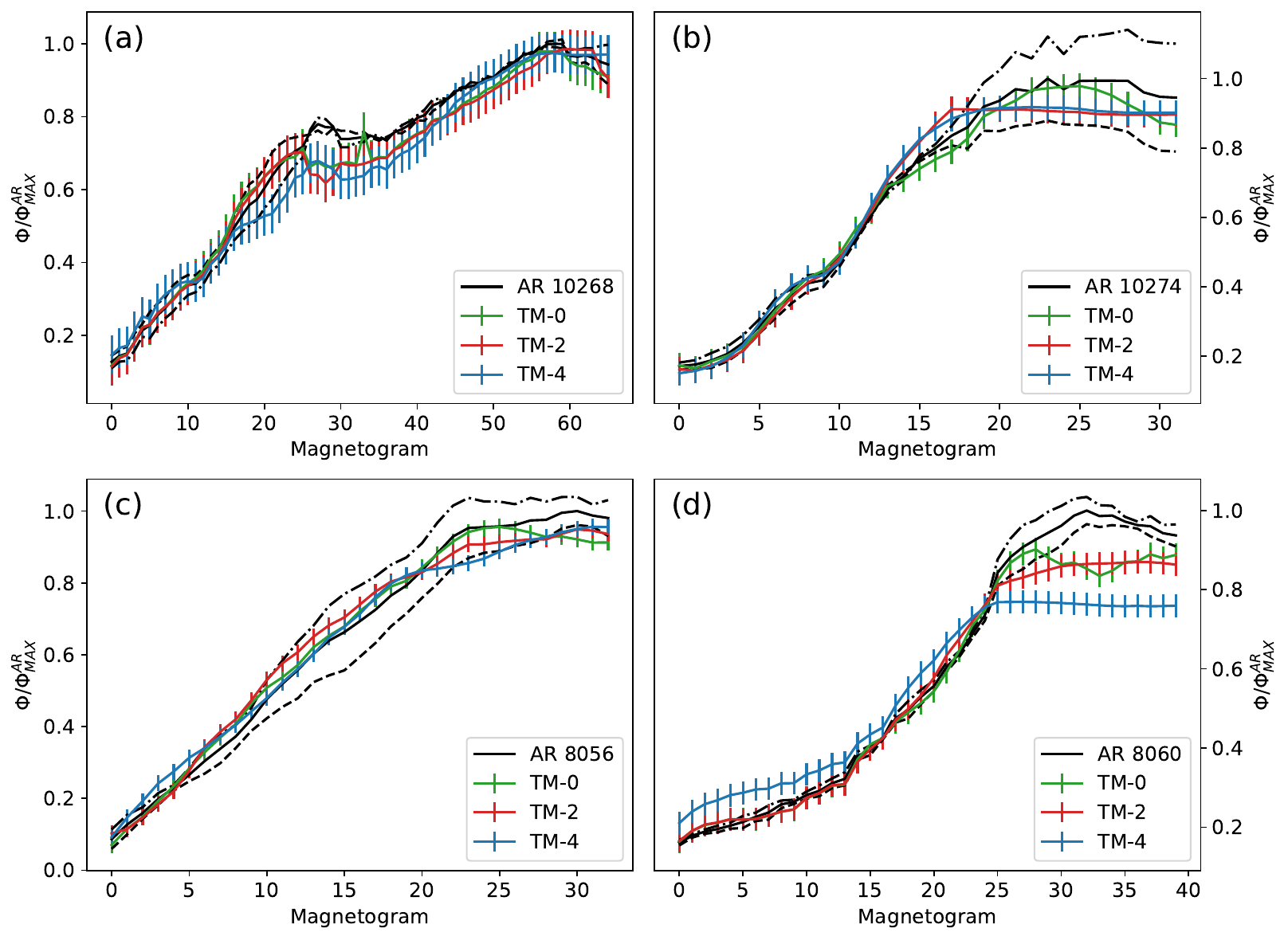}}
 \caption{Evolution of the observed and modeled photospheric magnetic fluxes for ARs 10268 \textbf{(a)}, 10274 \textbf{(b)}, 8056 \textbf{(c)}, and 8060 \textbf{(d)}. All curves are normalized with the maximum observed flux $\Phi_{MAX}^{AR}$. \textit{Black lines} show the evolution of the net observed magnetic flux. \textit{Dashed} and \textit{dash-dotted lines} correspond to the positive and negative fluxes, respectively. The flux derived from the models uses the most probable parameters. The flux uses the same color convention as in \fig{AR-params1}. Error bars for the modeled fluxes are computed using the same standard error per pixel as for the inference (\sect{inference}).
} \label{fig:AR-flux}
\end{figure}

\fig{AR-flux} shows the evolution of the observed and the modeled photospheric fluxes for the analyzed ARs. 
The black curves correspond to the net AR magnetic flux (solid line), the positive flux (dashed line), and the negative flux (dash-dotted line), all normalized with the AR maximum net flux $\Phi_{MAX}^{AR}$. 
Colored curves show the evolution of the modeled flux using the TM-0 (green), the TM-2 (red), and the TM-4 (blue), also normalized with the same factor $\Phi_{MAX}^{AR}$. 
Modeled fluxes are computed using the parameters obtained from the median of the posteriors. 
Error bars correspond to the flux uncertainties propagated from a standard deviation of the magnetic field equal to $\sigma$ per pixel.
All most probable models reproduce most of the AR flux evolution, indicating that a large percentage of the AR flux can be modeled with our current TMs.
The largest departure between the modeled and the observed flux is found at the end of the emergence, especially for ARs 10274 (panel b) and 8060 (panel d). 
This is due to the dispersion of the polarities at those times, especially in cases in which the dispersion increases the asymmetry between leading and following polarities so that the FR model becomes less appropriate to model the observed magnetograms. {Examples of the model magnetograms obtained with TM-0, TM-3, and TM-4 can be found in the supplementary movie files for each of the analyzed ARs (10268-mov.mp4, 10274-mov.mp4, 8056-mov.mp4, and 8060-mov.mp4).}

\begin{figure} 
 \centerline{\includegraphics[width=0.95\textwidth,clip=]{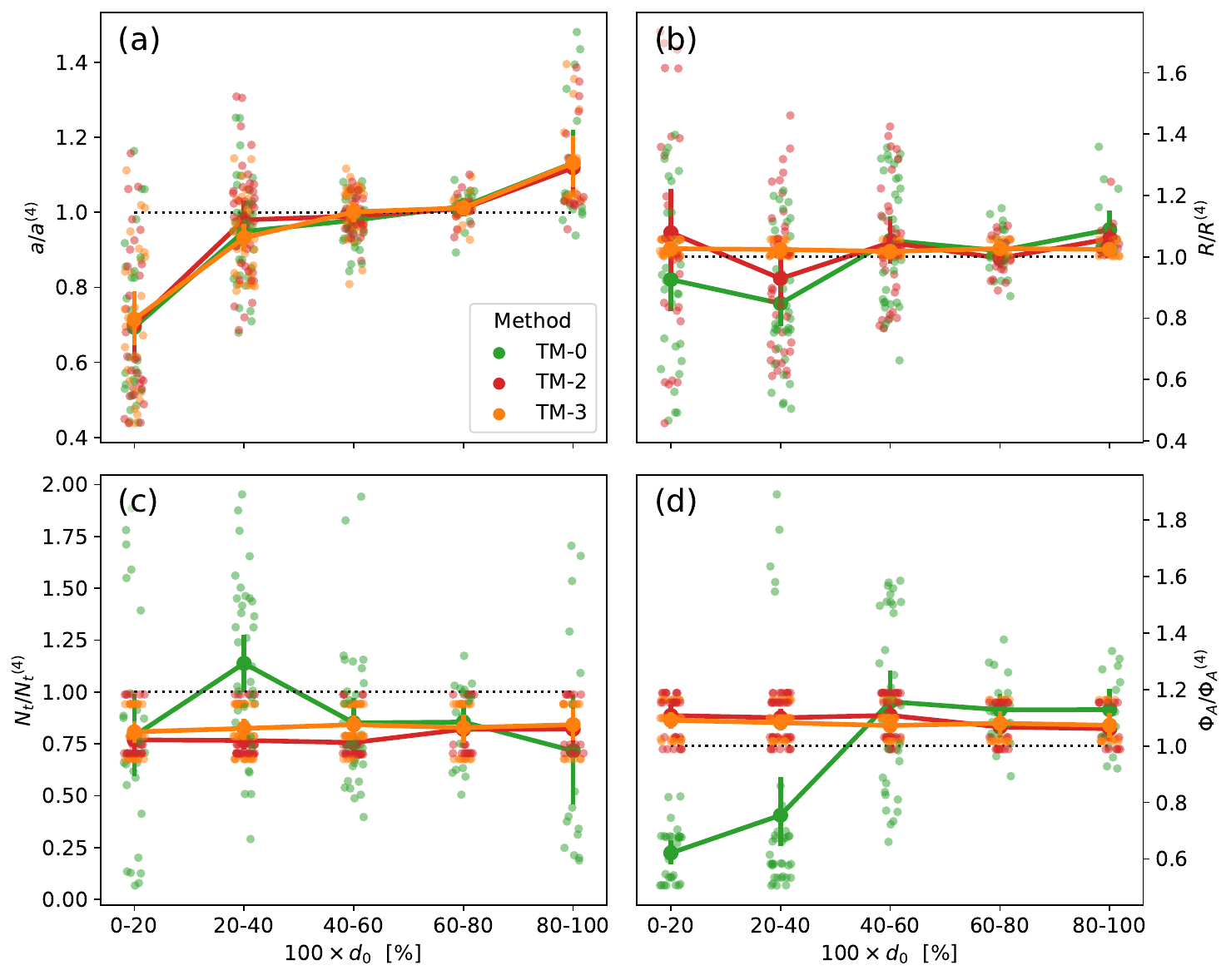}}
 \caption{\textbf{(a)} to \textbf{(d)} show the ratio between the inferred field parameters ($a$, $R$, $\Nt$, and $\Phi_A$) for different TMs to the respective value obtained with TM-4 as a function of the fraction of the emerged FR obtained with TM-4. The five bins include the parameters obtained with TM-0 (\textit{green}), TM-2 (\textit{red}), and TM-3 (\textit{orange}) for ARs 10268, 10274, 8056, and 8060. The \textit{large dot} and \textit{vertical lines} correspond to the median and quartile over each bin. \textit{Dashed black horizontal lines} correspond to values obtained with TM-4 (equal to one due to the normalization).
} \label{fig:AR-params2}
\end{figure}

We next analyze the behavior of the field parameters, which are $a$, $R$, $\Nt$, and $\Phi_A$. {In order to decrease the number of figures}   
we provide a summary by grouping the AR results.   This requires the definition of the same abscissa which is {comparable} 
to a normalized time.  We selected $d_0$ for a common abscissa since $d_0$ indicates the fraction of the FR located above the photosphere (\eq{dnot}), and is a more physical quantity to represent the four AR evolutions. Since TM-4 has a monotonous temporal evolution for all ARs (\fig{AR-params1}), we select $d_0$ estimated with TM-4 for the abscissa.  Then, since the detailed evolution of each AR is not relevant, we define time bins for $d_0$. We group the magnetograms of the ARs within five bins defined by the percentage fraction of the emerged FR. Finally, to compare the parameter results, in \fig{AR-params2} we plot the ratio of the inferred field parameters, $a$ (panel a), $R$ (panel b), $\Nt$ (panel c), and $\Phi_A$ (panel d), obtained with TM-0, TM-2, and TM-3 to the corresponding TM-4 values. Colored points indicate each TM (same color convention used in \fig{AR-params1}) for the four ARs. Large dots and vertical lines mark the median and quartile over each bin. Solid lines represent the evolution of the median. The dashed horizontal black line indicates the values for which the parameters are equal to those obtained with TM-4.

 If we consider the TM-4 as the reference model, then we find some departures of the estimated parameters using the other three TMs. \fig{AR-params2}a shows that the FR radius, $a$, can be consistently estimated when the FR has emerged above {$d_0=0.2$.} 
 This estimation of $a$ is independent of the TM used, therefore analyzing single magnetograms with TM-0 also provides a good approximation of this parameter (except for the first bin). It also shows that the radius $a$ is growing with $d_0$ for all TMs where this parameter is free to evolve (for TM-0 to TM-3).
 
 Similarly, the parameter $R$, shown in \fig{AR-params2}b, coincides for all TMs with the estimation of TM-4 at all the different stages of the AR emergences. Therefore, the estimations of $R$ are consistent between any of the proposed TMs.
 
{$\Nt$ and $\Phi_A$ determined with TM-0 (green lines) have the largest shift at the beginning of the emergence, as expected since they have no constraint so they are not well determined when only a little part of the FR has emerged.} 
The 
values for parameters $\Nt$ and $\Phi_A$ are not strictly constant for TM-3 and TM-2 because of statistical fluctuations within each of the bins (\fig{AR-params2}c-d). These parameters are slightly shifted from the reference TM-4.  For TM-3 and TM-2, the number of turns is $0.75$ times the values estimated with TM-4, and the axial flux is around $1.1$ times this reference. This opposite departure for these two parameters suggests that complex correlations between parameters might be present in the TMs. These correlations imply that the observed tongues can be modeled with a different combination of parameters depending on the selected TM. In the next section, we analyze how these correlations affect the estimation of the parameters.

\subsection{Comparison of Mean Parameter Values}

{Below we provide} a comparison of the number of turns, $\Nt$, and tilt $\alpha$ between TM-2, TM-3, TM-4 and previous estimations. Table~\ref{table2} compares the estimated values of $\Nt$ for the TMs and the method developed in \citet{Poisson16}, where an angle called $\tau$ is computed, {then we use $\tau$ as a label for this method in Table~\ref{table2}}. 
Since the magnetic tongues affect the inclination of the PIL, the angle $\tau$, defined as the acute angle between the PIL and the bipole axis is related to the twist of the FR. The half torus model provides an approximation for $\Nt$ using the mean estimated $\tau$ along the AR emergence. 
This $\Nt$ is a mean value that depends on how long and how much the tongues affect the PIL inclination. In general, as it was also tested in \citet{Poisson15a}, this estimation provides a lower bound for the value of $|\Nt|$. 

The results in Table~\ref{table2} show that the twist sign is the same for all TMs for each analyzed AR. For all ARs, TM-4 provides the largest values of $|\Nt|$. 
Indeed, the variation of $a$ and $R$ in the other TMs (TM-0 and TM-2) have also an impact on the shape and elongation of the tongues. In particular, the largest differences of $\Nt$ between TM-4 and the other two TMs correspond to AR 8060. This AR also presents a different evolution of the tilt given by TM-4 compared to the other two (see \fig{AR-params1}d-2) suggesting that the evolution of $a$ is significant in this case.  Imposing it as a constant for TM-4 implies a constraint with a strong impact on the inferred tilt.

  \begin{table} 
\caption{Estimated parameters for the number of turns ($N_t$) and the mean tilt ($\bar{\alpha}$) for ARs 10268, 10274, 8056, and 8060. Column ``lat" indicates the heliographic latitude of the AR emergence. $N_t$ is estimated as the median of its posterior distribution obtained with the TMs, $\tau$ indicates the estimation made with the PIL inclination {on the bipole axis} in \citet{Poisson16}. The mean tilt $\bar{\alpha}$ is computed from the parameter $\alpha$ along the AR evolution between $t_c$ {(see dashed lines in \fig{AR-params1})} and the last available magnetogram. The column ``Bar." refers to the estimation of the tilt obtained with the magnetic barycenters.
} \label{table2}
\label{T-complex}
\begin{tabular}{lccccccccc} 
  \hline
AR & lat. [deg] & \multicolumn{4}{c}{$N_t$
}  
     & \multicolumn{4}{c}{$\bar{\alpha}$ [deg]
     } 
     \\
     &      &   TM-4 & TM-3 & TM-2 & $\tau$ & TM-4 & TM-3 & TM-2 & Bar. \\
  \hline
10268 & 12 & -0.73 & -0.68 & -0.55 & -0.86 & -20  & -20 & -17 & 3  \\
10274 & -7 & 0.96 & 0.90 & 0.95 & 0.61 & -14  & -15 & -14& -30  \\
8056 & 17 & -0.50 & -0.39 & -0.35 & -0.29 & -23  & -22& -22& -16  \\
8060 & 5 & 0.91 & 0.61 & 0.64 & 0.20 & 13  & 11 & 12 & 1  \\
  \hline
\end{tabular}
\end{table}

The mean tilt $\bar{\alpha}$ is computed as the temporal mean of the tilt over the time range defined between the time $t_c$ {(dashed lines in \fig{AR-params1})} and the last available magnetogram for each AR. $\bar{\alpha}$ values are consistent between the TMs with standard deviations below $3^\circ$, indicating that the tilt converges to similar values for each AR within the estimated uncertainties {of TR methods}. The differences between the tilt estimated with the TMs and the magnetic barycenters are important, especially for ARs 10268, 10274, and 8060, and it scales with the value of $|\Nt|$. This difference is due to the effect of strong tongues on the determination of the barycenters, especially when the tongues persist until the last part of the emergence.

\section{Model comparisons} \label{s:compare} 

In the previous section, we obtained an estimation of the magnetic parameters of emerging ARs using different TMs. Some of these parameters are consistent between the different methods, but others show a significant deviation (see for example \fig{AR-params2}b-d). We recall that the evolution of the $d_0$ parameter for TM-0 does not correspond to the continuous upward emergence of a single FR, so TM-0 results appear unphysical. However, there is no indication that any of the different solutions found with the other TMs are more physically consistent. In this section, we compare the inference done with all the TMs, analyzing the corresponding errors, and the correlation between the parameters, and testing the stability of the posterior when there is missing information from the original data.

\subsection{Standard Deviation of the Magnetograms}

First, we compare the modeled data with the observations using the mean standard deviation 
($\sigma_{\rm est}$) as defined in \citet{Poisson22}:

\BE \label{eq_sigmaEstim}
\sigma_{\rm est} =  \sqrt{ \frac{1}{N_d} \sum_{j=1,N_d} (\Bj{o}-B_{\rm TM,j})^2 } \,,
\EE

\noindent where $N_d$ is the number of pixels considered in the summation, 
$\Bj{o}$ and $B_{\rm TM,j}$ are the observed and modeled field strength of pixel $j$ for a single magnetogram. {The computations of $\sigma_{\rm est}$ is done independently for each magnetogram along the AR evolution.}
The index TM represents any of the temporal methods (TM-0 to TM-4). The most probable results are used to compute $B_{\rm TM,j}$.
Pixels index $j$ correspond only to those in the mask defined for each magnetogram.

\begin{figure} 
 \centerline{\includegraphics[width=0.95\textwidth,clip=]{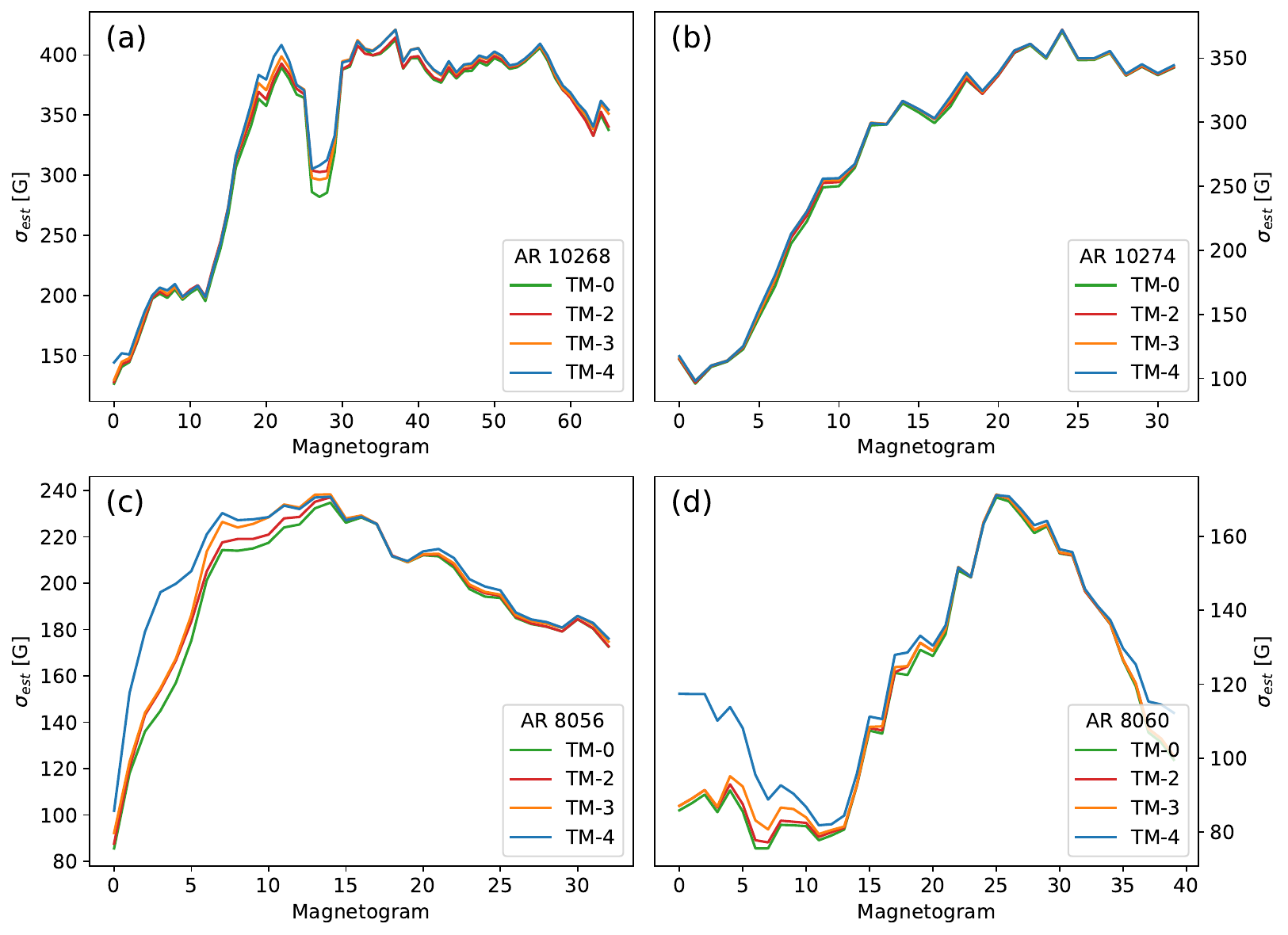}}
 \caption{Evolution of $\sigma_{\rm est}$ for ARs 10268 \textbf{(a)}, 10274 \textbf{(b)}, 8056 \textbf{(c)}, and 8060 (d) for all four different TMs (same color convention as \fig{AR-params1}). 
} \label{fig:estimS}
\end{figure}

\fig{estimS} shows the evolution of $\sigma_{\rm est}$ for each TM (same color convention as in \fig{AR-params1}). We find no significant differences between TMs for ARs 10268 (panel a) and 10274 (panel b). This indicates that all TMs provide similar field distributions despite their different most probable parameters. In all analyzed ARs, TM-0 provides the best fit (the lowest $\sigma _{\rm est}$ is obtained as expected since TM-0 has the largest number of free parameters), but in most cases, differences between the other TMs are below $4$ G. The largest differences correspond to the TM-4, as expected since it is the more constrained model. There is a significant increase in the errors at the early phases of the emergence of ARs 8056 (\fig{estimS}c) and 8060 (\fig{estimS}d). These results are consistent with the differences observed in the evolution of $\alpha$ in \fig{AR-params1}.

In general, the evolution of $\sigma_{\rm est}$ relates to the complexity of the AR flux since as the emergence becomes more complex the largest discrepancies with the model appear. Nevertheless, the peak of these deviations is affected also by the asymmetry of the polarities and the flux imbalance. This explains why the peak of $\sigma_{\rm est}$ is not always necessarily associated with the maximum flux of the AR.

\subsection{Correlation between parameters} \label{s:corr}

\begin{figure} 
 \centerline{\includegraphics[width=0.95\textwidth,clip=]{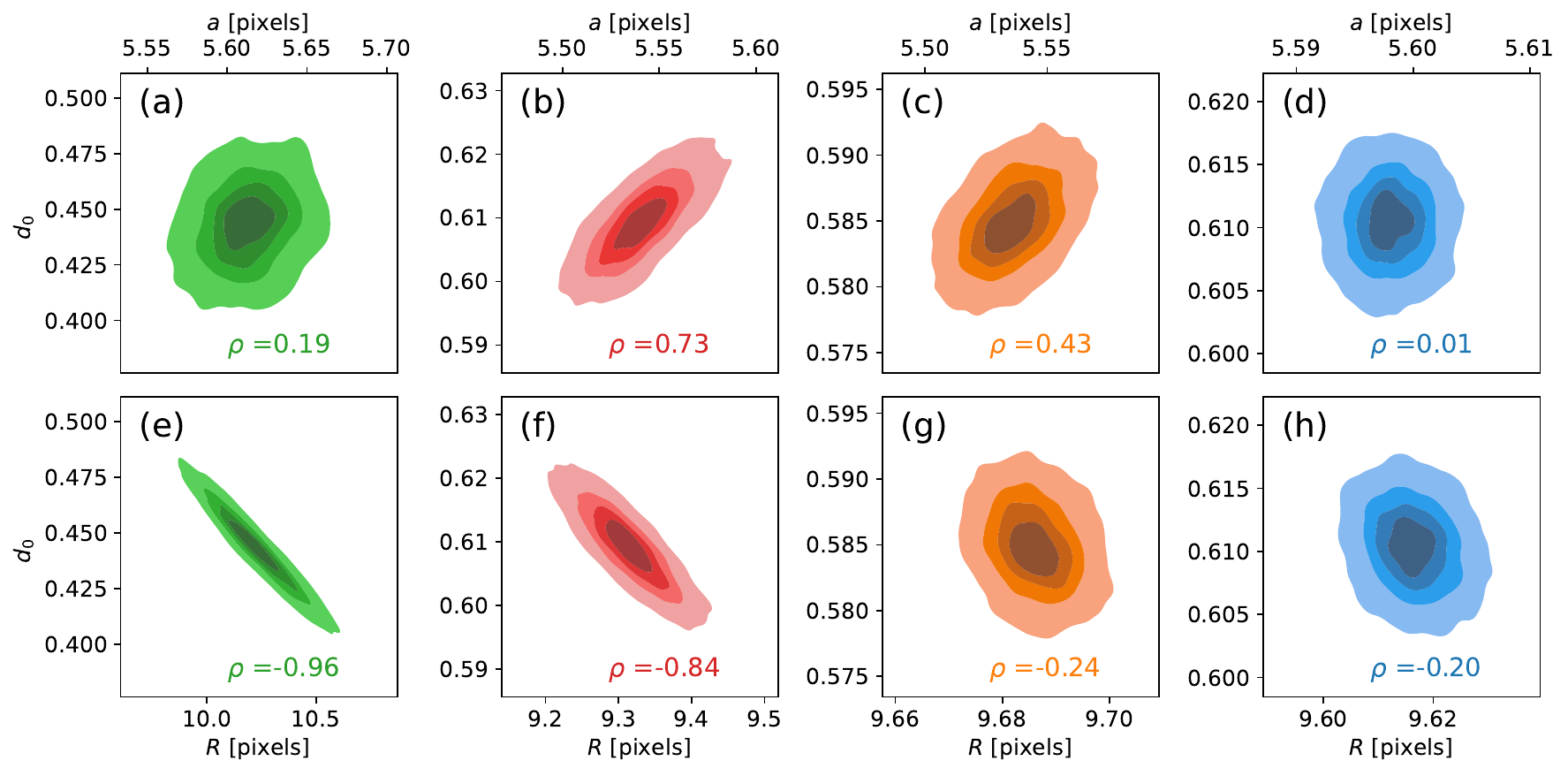}}
 \caption{Joint probability distributions obtained with TM-0 (\textit{green}), TM-1 (\textit{red}), TM-2 (\textit{orange}), and TM-4 (\textit{blue}) for parameters $a$, $R$, and $d_0$ of AR 10268 magnetogram number 50. Within each panel, we include the Pearson correlation coefficient $\rho$ between the joint distributions. 
} \label{fig:CORR-10268}
\end{figure}

How our model is constructed produces correlations between some of the parameters, {\eg}, $\Phi_A (a)$ or $d_0 (a,R)$. Other correlations might also appear simply because the magnetogram data are insufficient to constrain the FR model parameters completely (because the posterior probabilities are comparable for different sets of FR parameters). Strong correlations could impact the degeneracy level of the model. We interpret the degeneracy as the capability of the model to efficiently reproduce the same data with a different combination of parameters. Therefore, correlations between 2 parameters with Pearson's coefficients $\rho \approx 1$ (or $-1$ for anticorrelated parameters) imply that any solution with similar probability along the direct linear relationship existing between these parameters might be considered equally suitable to model the observations. \fig{CORR-10268} shows an example of joint probability distributions for $d_0$ with the parameters $a$ (top panels) and $R$ (bottom panels). These distributions correspond to the inferred parameters obtained from magnetogram 50 of AR 10268. Colors and panels represent the different TMs using the same color convention as defined in \fig{AR-params1}.

We found a significant increase in the correlation between the parameters in the case of the TM-0 and TM-2 with the absolute value of the correlation coefficients above $0.8$ for $d_0(R)$ (panels e and f). This strong correlation can produce multiple solutions for the FR model parameters for a single magnetogram (within a probability isocontour with an elliptical shape). In particular, this kind of correlation means that both parameters, $d_0$ and $R$, play equivalent roles in modeling the emergence. On the other hand, the most restrictive TMs, such as TM-3 and TM4, produce uncorrelated parameters (see orange and blue distributions in \fig{CORR-10268})

The other possibility in which degeneracy might be present is the existence of multimodal posterior distributions with several local maxima of the likelihood function. The chosen sampler is fundamental in detecting these cases, and it is common with Hamiltonian Monte Carlo algorithms (such as NUTS) to get stuck on a small region of the parameter space in which a local maximum of the likelihood is found. That is why we perform the sampling with four independent chains, each with different initial points scattered over the full parameter space. We look at the normalized rank $\hat{R}$, defined at the end of \sect{inference}, to determine if the posterior converges to a single distribution. We compute $\hat{R}$ in all our analyses and found that this parameter is always below $1.1$, indicating a good sampler convergence. 

Moreover, in \citet{Poisson22} we performed a test of the TM-0 using a Sequential Monte Carlo sampler \citep{KANTAS2009}, as an alternative to NUTS. It provides, with a larger computational demand, a more comprehensive sampling of the parameter space and is therefore more sensitive to multimodal distributions. For AR 10268 we found that this sampler also converged to a single normal distribution.

\subsection{Model Stability Test} \label{s:stab}

%
 \begin{figure} 
 \centerline{\includegraphics[width=0.95\textwidth,clip=]{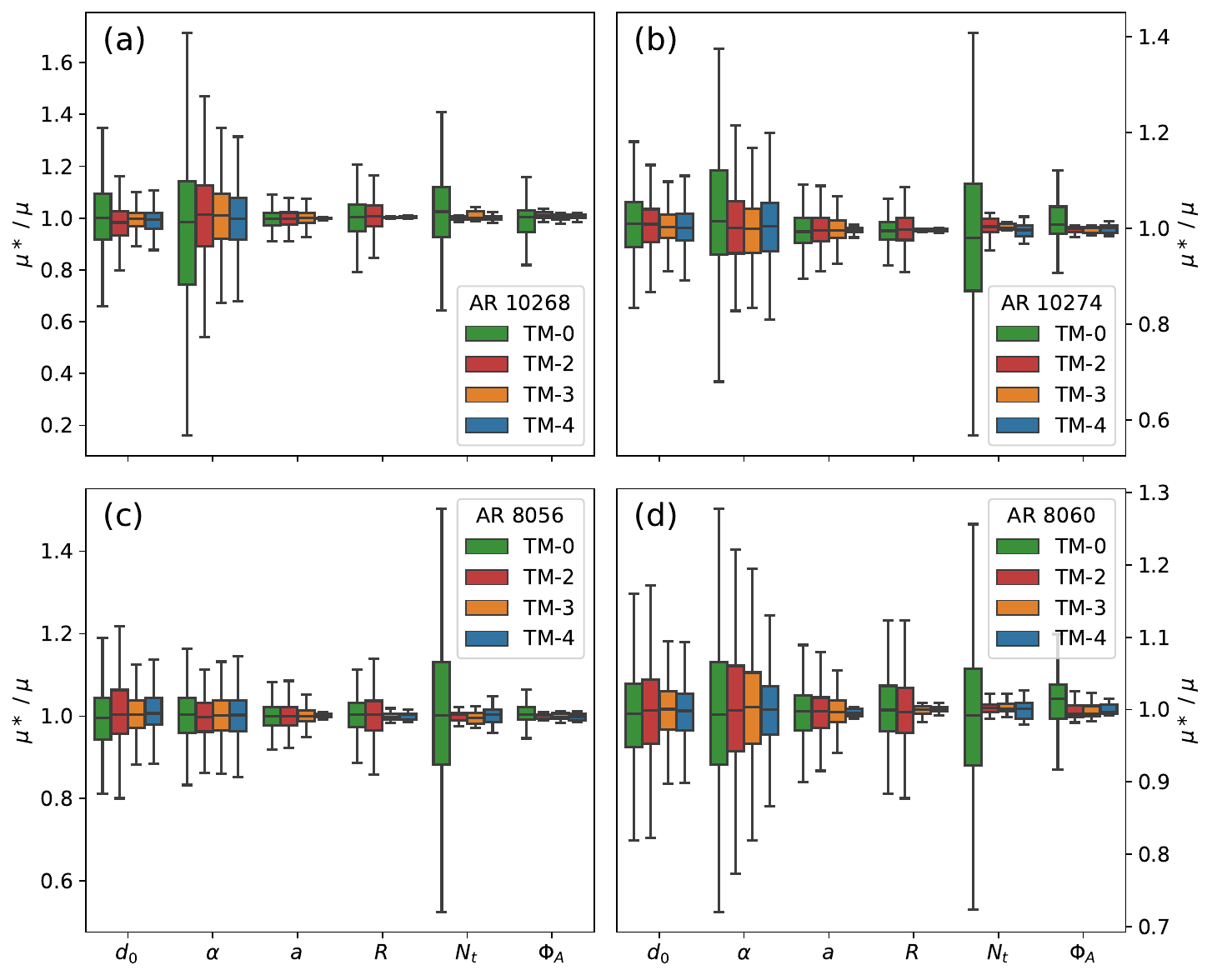}}
 \caption{Ratio between the 
 {parameter means} obtained with the altered data, $\mu^*$, and with original data $\mu$. Boxes and whiskers indicate the quartile and 1.5 of the interquartile range of these ratios for each parameter in $\vp$. The same color convention as in \fig{AR-params1} is used to identify each TM. \textbf{(a)} to \textbf{(d)} Correspond to ARs 10268, 10274, 8056, and 8060, respectively.
} \label{fig:stability}
\end{figure}

We conduct a stability test to ensure the accuracy and reliability of our TMs. The test involves the removal of 50\% of the pixels from the original data $\Mo$ and then re-computing the TMs with the altered data $M_a$. We conduct $12$ rounds of random selections of pixels from the original data cubes. Then we obtained an inferred posterior $P(\vp|M_a)$ for all the TMs. $P(\vp|M_a)$ is a distribution with the same dimension (8 parameters) and the same number of samples {(8000)} as the original posterior $P(\vp|\Mo)$, but with observations containing less information than the original data set.

To compare the difference between the inferred parameters $p$, we compute the ratio between the parameter mean values obtained with the altered data and with the original set. The means are computed as the mean of each marginal posterior, $\mu^* = \sum p\, P_m(p|M_a)$ for altered data or $\mu = \sum p\, P_m(p|\Mo)$ for the original data set. Since these distributions converged to Gaussian-like distributions the mean of the parameters is almost equivalent to the most probable value and the median for each parameter.

\fig{stability} shows the combined distributions of the ratio $\mu^*/\mu$ for each selection of points (ordinate) and each parameter (abscissa). The color boxes and whiskers represent the quartile and $1.5$ of the interquartile range, respectively. Panels correspond to ARs 10268 (a), 10274 (b), 8056 (c), and 8060 (d). Each distribution contains at least $12$ values for the ratio corresponding to each of the selected points of the altered data cubes. {More precisely for constant parameters, it contains $12$ values while for evolving parameters, this number is multiplied by the number of magnetograms studied} along the AR evolution. 

If the distributions are shifted from unity this implies that the mean of the parameters systematically changes when the data information is reduced (not detected in these cases). The increase in the broadness of the distributions implies that mean values are affected by reducing the input data.
In other words, the broadness of these distributions indicates how stable are certain parameters and how well they are constrained by the data. Extended distributions also indicate that the parameter space is probably too large and oversampling is occurring. We find that the TM-0 parameters are the most affected by the modified sampling (indicated by the green boxes). Among these parameters, $\Nt$ is the most sensitive for TM-0, since it is the only method that allows for its temporal evolution. In contrast, the axial flux ($\Phi_A$), which is also allowed to evolve for TM-0, is comparatively more stable.

 The parameters $a$ and $R$ are the most stable for all TMs. Combining this information with the results shown in \fig{AR-params2}, we conclude that these parameters are well constrained by the observations, and they are comparatively consistent between the TMs. For the other parameters, we see that using more restricted TMs results in more stable outcomes, as expected for a smaller parameter space. In this regard, TM-3 and TM-4 are more stable. However, we observe that $\alpha$ is one of the parameters that is most affected due to the loss of information. This indicates that the behavior of $\alpha$ is strongly influenced by the distribution of flux in each magnetogram. This means that decreasing the cadence and/or the spatial resolution of the data results in a less well-defined parameter.

\section{Summary of Results} \label{s:summary} 

Bayesian methods provide a robust framework for analyzing limited and uncertain data. This approach is particularly useful when dealing with sparse and noisy measurements, providing better uncertainty characterizations and enabling the development of more descriptive models. In this article, we develop Bayesian temporal methods to model LOS magnetograms of emerging ARs. Our goal is to constrain the physical parameters of the FR that originates the AR. In this regard, Bayesian techniques prove to be an important tool for improving the overall understanding of intricate flux emergence processes.

We use a FR emergence model, which is based on a half-torus magnetic field structure defined by eight parameters, to model the observed magnetograms of four bipolar ARs. The torus parameters that we infer include  $a$ (small radius), $R$ (large radius), $\Nt$ (number of turns of magnetic field lines around the FR axis or twist number), $\Phi_A$ (axial flux),  $d_0$ (fraction of the FR above the photosphere, \eq{dnot}), $\alpha$ (tilt angle), {$x_c$ and $y_c$ (horizontal coordinates of the FR center)}. To improve the method used in \citet{Poisson22} (defined here as TM-0), we introduce a temporal correlation of the parameters. We define temporal methods (TM) based on the number of parameters set constant along the AR evolution. TM-0 is the model with the largest parameter space, in which all parameters can evolve along the temporal dimension of the data. TM-4, on the other end, sets as constant {during the AR emergence} the parameters $a$, $R$, $\Nt$, and $\Phi_A$. In between these two extreme cases, TM-2 sets $\Nt$ and $\Phi_A$ as constant, while TM-3 sets $R$, $\Nt$ and $\Phi_A$ as constant during AR emergence.

According to the half-torus model, the emergence of the ARs relates to the increase of the parameter $d_0$ from 0 to 1. TM-0 is unsuccessful in describing a proper emergence, as the inferred evolution of $d_0$ is not consistent with the continuously upward emergence of a single coherent FR, as seen in the left panels of  \fig{AR-params1}. TM-2 improves the evolution of $d_0$, but some fluctuations appear due to the strong correlation between $d_0$ and $R$ (this correlation is analyzed in \sect{corr}). 
Among all the methods,  TM-3 and TM-4 present a monotonic increase of $d_0$ that best correlates with the increase in AR flux (see left panels in \fig{AR-params1} and \fig{AR-flux}).
Moreover, all the studied ARs reach maximum values of $d_0$ below unity despite the maximum flux being reached. This may indicate that a stage of emergence is reached in which the FR stops being buoyant. 

The tilt angle estimation obtained from the FR model removes the effect of magnetic tongues, which leads to a more precise estimation of the FR intrinsic tilt compared to those based on the computation of the magnetic barycenters (as seen in the right panels of \fig{AR-params1}). While the TMs generally show consistent results, in some cases, the values of the tilt angle, $\alpha$, obtained with TM-4 are found to be different from other TMs, particularly in the first part of the evolution of ARs 8056 and 8060 (as shown in panels c-2 and d-2 in \fig{AR-params1}). However, this discrepancy is not present for TM-3, which {sets the radius $a$ as an evolving parameter during the AR emergence. This} indicates that the freedom to evolve the parameter $a$ plays a role in stabilizing the determined tilt $\alpha$. $\alpha$ fluctuates significantly in the initial stages of emergence but tends to converge to a determined value at a specific stage. This stage relates to the time of half emergence of the FR top cross-section. {Finally, apart from specific cases mentioned above, all TMs significantly improve the estimation of the tilt angle during AR emergences, contributing to a better characterization of its spatiotemporal variations, then potentially of its origin.}

Consistent estimations of $a$ and $R$ are found for all TMs (see \fig{AR-params2}), except for $a$ derived by TM-4 at the beginning of the emergence. This means that these parameters are well constrained by the observations. { TM-2 and TM-3 present similar and consistent values of $\Nt$ and $\Phi_A$. In particular, the twist sign is in agreement with previous estimations made in \citet{Poisson15a}. We find an overestimation of $\Nt$ obtained with TM-4 in three of the four analyzed ARs. This indicates that the evolution of the parameter $a$, as in TM-2 and TM-3, also contributes to the reproduction of the observed elongation of the AR polarities. We {end up with} 
significant differences between the $\Nt$ obtained with the Bayesian method and the $\tau$ method from \citet{Poisson15a}.}

We find that the modeled magnetogram standard error ($\sigma_{\rm est}$) of the most probable FR model is consistent for all methods. Only TM-4 shows significant differences in some {ARs at the very early beginning of emergence} 
(see \fig{estimS}). This means that TM-0, TM-2, and TM-3 model the same observations with similar standard errors despite their differences in the inferred parameters. 

The correlation test, analyzed in \sect{corr}, shows that TM-0 and TM-2 present the strongest correlation between the parameters $a$, $R$, and $d_0$. This indicates that the parameter space is too large and the effect of modifying some parameters generates equivalent models.
On the contrary, more constrained TMs, TM-3 and TM-4, remove most of these correlations. Finally, the stability test presented in \sect{stab} shows that TM-0 presents the largest variations, meaning that over-fitting of the observations is occurring. Comparatively, the tilt $\alpha$ is the parameter that is more sensitive to the loss of data information in all TMs. This indicates difficulties in the estimation of this parameter for noisy or incomplete data sets.

\section{Conclusion} \label{s:conclusions} 

This research highlights the effectiveness of Bayesian methods in characterizing the behavior of emerging ARs. In particular, we conclude that TM-3 ({with imposed} constant $R$, $\Nt$, and $\Phi_A$ {during the full emergence}) presents the best performance as summarized as follows. TM-3 evolution of $d_0$ {is coherent with the monotonous upward emergence} 
of a single FR for all analyzed ARs (see left panels in \fig{AR-params1}). $\alpha$ obtained with TM-3 is {always consistent with TM-0 and TM-2} 
(see right panels in \fig{AR-params1}), and it corrects the spurious rotation obtained with TM-4 for AR 8056. The standard error of TM-3 is comparable with that of methods with larger parameter spaces, TM-2 and TM-0, implying that TM-3 effectively reproduces the observations (see \fig{estimS}). The general low correlation coefficient between $d_0$, $R$, and $a$ suggests that the sampled parameter space is not excessively large to lead to over-fitting of observations. 

{The methods presented in this work provide a consistent estimation of the magnetic parameters of ARs at the early stages of their evolution which may contribute to a better understanding of the origin of the emerging FR.} However, it is important to note that TMs with strong constraints, such as TM-4, can introduce biased results. Conversely, those with large and unconstrained parameter spaces, such as TM-0, can lead to over-fitting of observations, which affects the stability of the solution. 

Addressing the aforementioned issues involves enhancing the accuracy of results through the incorporation of more descriptive models. By including parameters that correspond to additional observed properties, we can effectively reduce the standard error of the model. However, the possible caveats of excessively enlarging the parameter space have to be carefully analyzed due to the limited information provided by the observations. In this way, it is also necessary to incorporate further temporal information in the parameters to properly constrain the priors.

%
\begin{acks} 
{We thank the referee for his constructive suggestions that have improved the manuscript.}
The authors M.P., M.L.P., C.H.M., and F.G. are members of the Carrera del Investigador Cient\'{\i}fico of the Consejo Nacional de Investigaciones Cient\'{\i}ficas y T\'ecnicas (CONICET) of Argentina.  
The authors acknowledge the use of data from the MDI (NASA) mission. 
We recognize the collaborative and open nature of knowledge creation and dissemination under the control of the academic community, as expressed by Camille No\^{u}s at http://www.cogitamus.fr/indexen.html.
\end{acks}

 \begin{authorcontribution}
M.P. contributed to the data reduction, code development, and computation of models. All authors contributed to defining the used methodology. All authors contributed to the writing and reviewing of the manuscript.
 \end{authorcontribution}
 \begin{fundinginformation}
The authors M.P., M.L.P., and C.H.M. acknowledge financial support from the Argentinian grants PICT 2020-03214 (ANPCyT) and PIP 11220200100985 (CONICET).
 \end{fundinginformation}
 \begin{dataavailability}
The data used in this work are available from: \\
\url{http://jsoc.stanford.edu/MDI/MDI_Magnetograms.html}.
\end{dataavailability}
%
\begin{conflict}
The author C.H.M. is Editor-in-Chief of the journal Solar Physics; the article underwent a standard single-blind peer review process. The authors declare that they have no other conflicts of interest.
\end{conflict}

%

%

%

%
%

%
%
 \bibliographystyle{spr-mp-sola}
 \bibliography{biblio3}  
%
%
%
%

\end{article} 
\end{document}